\documentclass[a4paper,fleqn,usenatbib]{mnras}
\pdfoutput=1 

\usepackage{newtxtext,newtxmath}

\usepackage[T1]{fontenc}
\usepackage{ae,aecompl}

\usepackage{graphicx}	
\usepackage{amsmath}	
\usepackage{amssymb}	

\usepackage{xcolor}
\usepackage{multirow} 
\usepackage{pdflscape} 
\usepackage[inline]{enumitem} 

\newcommand{\degree}{\ensuremath{^{\circ}}} 

\graphicspath{{figures/}}

\title[Interpretation of disc wind diagnostics]{The interpretation of protoplanetary disc wind diagnostic lines from X-ray photoevaporation and analytical MHD models}

\author[Weber et al.]{
Michael L. Weber,$^{1}$\thanks{E-mail: mweber@usm.lmu.de (MLW)}
Barbara Ercolano,$^{1,2}$
Giovanni Picogna,$^{1}$
Lee Hartmann$^{3}$
\newauthor
Peter J. Rodenkirch$^{4}$
\\
$^{1}$Universit\"ats-Sternwarte, Ludwig-Maximilians-Universit\"at M\"unchen, Scheinerstr. 1, 81679 M\"unchen, Germany \\
$^{2}$Excellence Cluster Origin and Structure of the Universe, Boltzmannstr. 2, 85748 Garching bei M\"unchen, Germany \\
$^3$Department of Astronomy, University of Michigan, 500 Church Street, Ann Arbor, MI 48105, USA \\
$^4$ Institute for Theoretical Astrophysics, Zentrum f\"ur Astronomie, Heidelberg University, Albert Ueberle Str. 2, 69120 Heidelberg, Germany \\
}

\date{Accepted XXX. Received YYY; in original form ZZZ}

\pubyear{2019}

\begin{document}
\label{firstpage}
\pagerange{\pageref{firstpage}--\pageref{lastpage}}
\maketitle

\begin{abstract}
High resolution spectra of typical wind diagnostics ([OI] 6300~\AA ~and other forbidden emission lines) can often be decomposed into multiple components: high-velocity components with blueshifts up to several 100 km/s are usually attributed to fast jets, while narrow (NLVC) and broad (BLVC) low-velocity components are believed to trace slower disc winds. Under the assumption that the line-broadening is dominated by Keplerian rotation, several studies have found that the BLVCs should trace gas launched between 0.05 and 0.5 au and correlations between the properties of BLVCs and NLVCs have been interpreted as evidence for the emission tracing an extended MHD wind and not a photoevaporative wind.
We calculated synthetic line profiles obtained from detailed photoionisation calculations of an X-ray photoevaporation model and a simple MHD wind model and analyzed the emission regions of different diagnostic lines and the resulting spectral profiles. The photoevaporation model reproduces most of the observed NLVCs but not the BLVCs or HVCs. The MHD model is able to reproduce all components but produces Keplerian double peaks at average inclinations that are rarely observed. The combination of MHD and photoevaporative winds could solve this problem. Our results suggest that the Gaussian decomposition does not allow for a clear distinction of flux from different wind regions and that the line broadening is often dominated by the velocity gradient in the outflow rather than by Keplerian rotation. We show that observed correlations between BLVC and NLVC do not necessarily imply a common origin in an extended MHD wind.
\end{abstract}


\defcitealias{Ercolano2010}{EO10}
\defcitealias{Ercolano2016}{EO16}
\defcitealias{Blandford1982}{BP82}
\defcitealias{Banzatti2018}{B+19}


\section{Introduction}\label{sec:introduction}
The evolution and dispersal of protoplanetary discs is a crucial aspect for our understanding of planet formation. Observations have shown that discs typically disperse within a few Myr \citep{HaischJr.2001, Mamajek2009, Fedele2010, Ribas2014} and various mechanisms have been proposed to be driving or contributing to this dispersal: viscous accretion \citep{Hartmann1998}, planet formation and outflows in the form of jets and winds \citep[e.g. reviews by][]{Armitage2011, Clarke2011, Alexander2014, Ercolano2017a}. Such outflows can be launched over a large range of disc radii and by different processes: Radiation from the central star is able to heat disc material until it becomes unbound and escapes in a slow-moving thermal "photoevaporative" wind beyond as close as $\sim$1~au out to tens of au from the star, depending on the type of radiation (EUV, FUV, X-ray) \citep[e.g. reviews by][]{Armitage2011, Alexander2014, Ercolano2017a}. Fast-moving winds and jets are usually attributed to MHD processes, which are also capable of launching winds very close to the star, where escape velocities are too high for a thermal wind to be driven by the central star. Recent theoretical simulations show that MHD winds can be launched over a large range of disc radii where they can remove disc material and angular momentum and thus drive accretion \citep{Gressel2015, Bai2016, Rodenkirch2019a}. However, these MHD models depend strongly on parameters that cannot be at present observationally constrained. The relative importance of photoevaporative winds and MHD driven winds is thus currently uncertain \citep[e.g.][]{Rodenkirch2019a, Kunitomo2020, Ercolano2017a}. Disc winds have been detected by observing small (order 1 km/s) blueshifts in emission lines from T-Tauri stars \citep[e.g.][]{Hartigan1995, Pascucci2009}. Determining the nature of these winds, whether thermally or MHD driven, is crucial to understand angular momentum transport in discs. Thermal winds do not transport angular momentum, while MHD winds do. Thus the mass-loss rates of putative MHD-driven winds can inform on whether they can account for the observed accretion rates in T-Tauri stars. To this aim it is important to compare theoretically predicted wind structures and kinematics with observations. 

Optical forbidden emission lines from oxygen and other atomic and lowly ionized elements such as sulfur can be used to trace outflows at least partially and their line ratios can provide information about the physical properties of the emitting region. The [OI]~6300~{\AA} line in particular has been subject to many studies and was often found to be blueshifted by a few km/s relative to the stellar velocity, indicating an origin in an outflow that is approaching the observer, while the emission from the receding outflow in the other half of the disc is obstructed by the dust disc \citep{Edwards1987}. The low resolution line profile can often be decomposed into a high-velocity component (HVC) with a blueshift of $\sim$30 up to a few hundred km/s and a low-velocity component (LVC) with lower blueshifts of a few km/s \citep{Hamann1994,Hartigan1995}. The HVC is usually attributed to collimated jets, evidence for which has been found in many observations \citep[e.g.][]{Ray2006, Dougados2000}. One of the best studied objects for which a jet has been observed is DGTau \citep[e.g.][]{Gudel2008, Guedel2011}.

\citet{Ercolano2010} simulated forbidden emission line profiles from a thermal wind that is driven by X-ray photoevaporation and found line luminosities that are in good agreement with observed LVCs. However, observational studies of high-resolution line profiles by  \citet{Rigliaco2013} and \citet{Simon2016a} showed that the LVC can often be decomposed further into a broad component (BLVC) with FWHM of at least $\sim$40~km/s and a narrower component (NLVC). \citet{Banzatti2018} further distinguish between low-velocity components that are composed of two components (BLVC and NLVC) and low-velocity components that are best fitted with a single component (SC). If the single component is accompanied by a high-velocity component they classify it as SCJ. While the typical width of the narrow component is reasonably well reproduced by X-ray photoevaporation models, the BLVC is too broad to have an origin in a photoevaporative outflow, if the broadening is dominantly due to Keplerian rotation. Instead, \citet{Simon2016a} suggested that the origin of the broad component could lie in an MHD-driven wind that is launched at radii of $\lesssim$ 0.5~au (vs. $\sim$0.5 - $\sim$5~au for the NLVC), thus closer than the launching radius of a photoevaporative wind and therefore much more subject to Keplerian broadening. Indeed they find the correlation between the FWHM of the BLVC and NLVC that is expected from Kepler-broadening, but \citet{McGinnis2018} point out a significant spread around that correlation as an indicator that other mechanisms contribute to the broadening as well. More evidence for physically distinct emission regions between the broad and narrow components is the [OI]~5577/6300 line ratio that is typically higher in the BLVC than the NLVC, as well as an observed correlation between the blueshift of the BLVC and the accretion luminosity that is absent in the NLVC \citep{Simon2016a}. Another correlation found in numerous studies is between the forbidden line luminosity and the accretion luminosity \citep{Rigliaco2013, Natta2014, Simon2016a, McGinnis2018, Fang2018, Banzatti2018}, which is well reproduced by \citet{Ercolano2016}.

Recently, \citet{Fang2018} analysed a large sample of high-resolution ($\sim$6~km/s) line profiles in order to extend the previous detailed studies of the [OI]~6300 line to the [OI]~5577 and [SII]~4068 lines and their line ratios. For the HVC, they found line ratios that are consistent with common shock models and based on the similarity of the [OI]~6300 and [SII]~4068 lines, which have similar critical densities and excitation temperatures, they rule out photodissociation as the dominating origin of [OI]~emission, in favor of a thermal origin in gas with T $\sim$5000 - 10000~K and n$_e$ $\sim10^7$ - $10^8$~cm$^{-3}$, in agreement with previous findings by \citet{Natta2014}. Most of their observed line ratios of the narrow components are compatible with a photoevaporative wind, but the FWHM of the broadest narrow components are larger than those predicted from the models of \citet{Ercolano2016}. In a parallel work, \citet{Banzatti2018} studied the kinematic behaviour of the individual [OI]~components using profiles from a large sample of 65 T-Tauri stars. Contrary to \citet{Simon2016a} they do find a positive correlation between the centroid velocities of the NLVCs and the accretion luminosity. They also find a correlation between the infrared-index and the SC properties, suggesting that the inner wind evolves together with the inner disc. Moreover they suggest that both, the broad and narrow low-velocity components, arise in the same MHD-driven wind based on tight correlations between BLVC and NLVC kinematics. A similar argument has been made by \citet{Nisini2017}, based on correlations of HVC and LVC properties with the accretion luminosity.

In this paper we calculate synthetic profiles of four forbidden emission lines using the improved X-ray driven photoevaporative wind model by \citet{Picogna2019} and the magnetocentrifugally driven wind model by \citet{Milliner2018}. We show that the observed correlations between NLVCs and BLVCs can indeed be reproduced, even when they have their origin in different wind types and that the commonly applied Gaussian decomposition of the line profiles is not necessarily suitable to analyze distinct physical regions of an outflow. Further, we show that hot shock regions in a collimated jet that illuminate and heat high layers of the winds in a "lamppost"-like fashion can affect the line profiles. In section \ref{sec:methods} we present the methods used in our study and discuss their limitations, in section \ref{sec:results} we present the main results, followed by a brief discussion in section \ref{sec:discussion} and in section \ref{sec:conclusions} we draw our main conclusions from this work.

\section{Methods}\label{sec:methods}
\subsection{X-ray photoevaporative wind model} \label{sec:met:model_pe}

A photoevaporative wind is launched when radiation heats up the gas in the disc and deposits enough energy to leave the gas gravitationally unbound. In the planet-forming region of a disc the heating process is generally dominated by irradiation from the central star. We use the velocity and density structure of the gas in a photoevaporating wind from the radiation-hydrodynamical calculations of \citet{Picogna2019}. We rerun the hydrodynamical calculation in order to extend the simulation domain to 1000 au above the disc midplane. This was necessary to account for lamppost illumination from a jet source. The inner and outer radius remained at 0.35 and 200 au as in \citet{Picogna2019}. In this model a primordial disc of mass 0.07 M$_{\odot}$ is irradiated by its 0.7 M$_{\odot}$ host star with the X-ray + EUV spectrum presented by \citet{Ercolano2008a, Ercolano2009} with a luminosity of $L_X = 2\cdot10^{30}$ erg/s. Solar abundances from \citet{Asplund2005}, depleted according to \citet{Savage1996} were assumed. The gas and dust radiative transfer code MOCASSIN \citep{Ercolano2003, Ercolano2005a, Ercolano2008} was used to obtain a parametrisation for the gas temperature as a function of ionisation parameter and gas column density. This parametrisation was then used in the hydro-code PLUTO to calculate the hydrodynamic structure of the wind. More details are given by \citet{Picogna2019}. 

We post-processed the density and velocity grids of the model by mapping it to a 2-dimensional Cartesian grid with 120 logarithmically spaced grid points in disc radius and 2000 linearly spaced points in height. The remapped grid extends from the disc radius R = 0 to 66 au and the height z = 0 to 200 au, however the inner radius of the hydrodynamical model is 0.35 au, below which all cells are set to 0. We have verified that the choice of 0.35 au for the inner radius is sufficient to accurately calculate profiles by running an additional hydrodynamical model with an inner radius of 0.05 au and comparing the resulting profiles. The emission from inside 0.35 au does not contribute significantly to the total flux, because the volume of this region is small. We choose the extent and resolution in z that high in order to be able to properly resolve the high regions of the wind when investigating the effect of lamppost illumination from a jet. The final grid results from averaging the last 25 orbits of the simulation in order to increase the smoothness. 

\subsection{Magnetocentrifugal MHD wind model} \label{sec:met:model_mhd}

A number of magneto-hydrodynamical calculations of discs have shown that under given assumptions for magnetic flux and ionisation level in the disc a wind is launched, which can be as vigorous as a photoevaporative wind and which could, in theory, extend to radii well within the gravitational radius for photoevaporation. Indeed \citet{Banzatti2018} (B19) suggest that an MHD wind might be responsible for the observed broad component of [OI] forbidden line emission. Unfortunately a post-processing of currently available MHD calculations is not suitable for this purpose, since the grids of the available calculations do not extend to small enough radii where the emission would have to originate to give the observed FWHM of the BLVC of the [OI] 6300 line.

We have indeed experimented by post-processing the recent calculation of \citet{Rodenkirch2019a} which include a MHD wind and a photoevaporative wind for a range of the plasma parameter $\beta$, i.e. the ratio of thermal pressure over magnetic pressure. Unfortunately similar to other calculations \citep[e.g.][]{Wang2019} their grid only extends to 1 au in inner radius, while the observed FWHM suggest emission regions $<$ 0.5 au. Furthermore a limitation of these calculations is the need to impose a floor density for the inner regions, which produces numerical artefacts that significantly affect the line profiles. The floor density is needed because with decreasing density the Alfv\'{e}n velocity, which has to be resolved in every dynamical time-step, increases, imposing a severe limit to the time-step which leads to an unfeasible increase of the computational time.

These are serious limitations for the calculations of line intensities and profiles originating from the inner regions of the computational domain, which preclude the use of these numerical calculations at present. 

In this work we thus rely on an analytical description of a magnetocentrifugal MHD wind. For a description of the density and velocity structure of the MHD wind we use the model by \citet{Milliner2018} that is based on the \citet{Blandford1982} (BP82) axisymmetric self-similar solutions for a magnetocentrifugally driven wind for thin discs. It was originally applied to FU Ori type objects, but is not specific to such. The model uses cylindrical coordinates [$R$, $\Phi$, $z$] = [$R_f\xi$, $\Phi$, $R_f\chi$] with non-dimensional functions [$\xi'$, $\Phi$, $\chi$] and the footpoint $R_f$ to describe self-similar streamlines of the form 
\begin{align}
    \chi = a\xi^2 + b\xi - (a + b).
\end{align}
\citet{Milliner2018} follow \citetalias{Blandford1982} by using the velocity components 
\begin{align}
    [v_R, ~v_{\Phi}, ~v_z] = [\xi' f(\chi), ~g(\chi), ~f(\chi)] \sqrt{GM_*/R_f},
\end{align}
to solve the structure of the flow along the streamlines and to finally derive the mass density
\begin{align}
    \rho v_z = (\rho v_z)(z = 0) \frac{(R_{in}/R_f)^2}{\xi(\xi - \chi \xi')}
\end{align}
with
\begin{align}
(\rho v_z)(z = 0) = \frac{\dot{M}_w}{4\pi R_{in}^2 \mathrm{ln}(R_{out}/R_{in})}.
\end{align}
where  $\dot{M}_w$ is the wind mass-loss rate and $R_{in}$ and $R_{out}$ are the inner and outer disc radii, respectively. More details can be found in \citet{Milliner2018} and \citetalias{Blandford1982}.

We use this model to construct 2-dimensional cartesian grids of the velocity and density structure with 100 grid points in disc radius and 600 points in height, which we found to be sufficient to properly resolve the emission. The grid points are spaced evenly in square root space. The grid extends from R = 0 to 10 au and z = 0 to 60 au. The parameters for the wind solution are $a = 0.0936$, $b = 0.612$, $\kappa = 0.03$, $\lambda = 30$, $R_{in} = 20$~$R_{\odot}$, $R_{out} = 10$ au and M$_*$ = 0.7 M$_{\odot}$. Since the parameters for MHD wind models are poorly constrained, we have chosen values that yield a poloidally-collimated flow similar to that seen in many numerical simulations. Based on the assumption that MHD winds drive accretion, we increase the mass-loss rates for models with a higher accretion luminosity to a rate comparable to the accretion rate \citep{Wang2019, Fang2018}. We choose $\dot{M}_w = 10^{-8.6}~\mathrm{M_{\odot}/yr}$ for models with an accretion luminosity of $2.6\cdot10^{-2}$~L$_{\odot}$, $\dot{M}_w = 10^{-7.5}~\mathrm{M_{\odot}/yr}$ for models with 0.31~L$_{\odot}$ and , $\dot{M}_w = 10^{-6.9}~\mathrm{M_{\odot}/yr}$ for 1~L$_{\odot}$. A summary of the input parameters for the different models is shown in table \ref{tab:input_parameters}. To calculate the hydrogen number density from the mass density we use the mean molecular weight $\mu = 1.3$. We assume the same abundances as in the X-ray photoevaporative model.

\begin{table*}
    \centering
    \begin{tabular}{ccccccccc}
    Model & $\mathrm{R_{in}}$ [au] & $\mathrm{R_{out}}$ [au] & $\mathrm{Z_{max}}$ & L$_{\mathrm{acc}}$ [L$_{\odot}$] & L$_{\mathrm{EUV}}$ [erg/s] & L$_{\mathrm{jet}}$ & log\big($\frac{\dot{\mathrm{M}}_{\mathrm{w}}}{\mathrm{M_{\odot}/yr}}$\big) \\
    \hline
    \hline
    PE-1     & 0.35 & 66 & 200 & $2.6\cdot10^{-2}$   & $8.56\cdot10^{28}$ & 0 & -7.8 \\
    PE-2     & 0.35 & 66 & 200 & 0.31                & $1.39\cdot10^{30}$ & 0 & -7.8 \\
    PE-3     & 0.35 & 66 & 200 & 1                   & $3.29\cdot10^{30}$ & 0 & -7.8 \\
    MHD-1    & 0.09 & 10 & 60  & $2.6\cdot10^{-2}$   & $8.56\cdot10^{28}$ & 0 & -8.6 \\
    MHD-2    & 0.09 & 10 & 60  & 0.31                & $1.39\cdot10^{30}$ & 0 & -7.5 \\
    MHD-3    & 0.09 & 10 & 60  & 1                   & $3.29\cdot10^{30}$ & 0 & -6.9 \\
    PEJ-1    & 0.35 & 66 & 200 & $2.6\cdot10^{-2}$   & $8.56\cdot10^{28}$ & 0.1~ L$_{\mathrm{acc}}$ & -7.8 \\
    PEJ-2    & 0.35 & 66 & 200 & 0.31                & $1.39\cdot10^{30}$ & 0.1~ L$_{\mathrm{acc}}$ & -7.8 \\
    PEJ-3    & 0.35 & 66 & 200 & 0.31                & $1.39\cdot10^{30}$ & 0.32~ L$_{\mathrm{acc}}$ & -7.8 \\
\end{tabular}
    \caption{Relevant parameters that were used to set up our models. PEJ models are photoevaporation models with an included "lamppost" illumination jet source.}
    \label{tab:input_parameters}
\end{table*}

\subsection{Photoionisation calculations} \label{sec:met:ionisation}

We follow the approach of \citet{Ercolano2010} (EO10) and  \citet{Ercolano2016} (EO16) to perform photoionisation calculations of the wind structures with the goal of predicting line luminosities and spectral profiles of the [OI]~6300, [OI]~5577, [SII]~4068 and [SII]~6730 lines. We use the previously mentioned MOCASSIN code with the same settings and input spectrum as in \citetalias{Ercolano2016} and \citet{Picogna2019}: A soft X-ray spectrum with a luminosity of $2\cdot10^{30}$ erg/s and an additional blackbody spectrum at a temperature of 12000~K that is loosely representative of an accretion component. We express the luminosity of the accretion component L$_{\mathrm{acc}}$ as the bolometric luminosity of the blackbody and perform our calculations for three different values: $2.6\cdot10^{-2}$~L$_{\odot}$, 0.31~L$_{\odot}$ and 1~L$_{\odot}$, corresponding to ionizing EUV luminosities (h$\nu$ > 13.6~eV) of $8.56\cdot10^{28}$~erg/s, $1.39\cdot10^{30}$~erg/s and $3.29\cdot10^{30}$~erg/s, respectively. Since the wind in our photoevaporative model is driven by the X-ray, the accretion component has no effect on the dynamics of this wind. For models where we investigate the effect of a "lamppost" illumination from a radiation source in the jet we add a third component, an EUV source modelled as a blackbody of temperature 25000~K. The bolometric luminosity of the component is chosen as a fraction of the accretion luminosity, where we calculate models with 10~\% and 32~\%. These parameters are observationally unconstrained and we use them because they work well to show the effect that a jet could have on the wind. The jet source is then placed on the z-axis at a height of 50 au. At that height, the wind velocity and density are approximately constant and a source higher in the jet would behave similarly.

\subsection{Line profile calculations and fits} \label{sec:met:profiles}

We have used the two-dimensional grids of emissivities from our photoionisation calculations and our grids of the wind structure to construct a three-dimensional axisymmetric cylindric volume and calculated the spectral profiles of our lines when viewed along different lines of sight. To construct the cylinder we used 256 evenly spaced azimuthal grid points. Based on the assumption that the disc midplane is optically thick to our lines we do not model the other half of the disc. This does have implications for our profiles at very high inclinations: Contribution from high regions in the receding side of the disc wind where the line of sight does not cross the midplane within the grid is lost in our calculation. However, we assume that this contribution is absorbed in the bound disc, because the line of sight does have a very long path through it, at the inclinations where this is relevant. We do account for line attenuation by dust absorption by calculating the column number density $N_H$ from each grid cell in the 3D-grid to the point where the line of sight leaves the grid. To calculate the absorption cross-section $C_{\lambda}^{abs}$ (5.49$\cdot10^{-23}$, 6.83$\cdot10^{-23}$, 1.52$\cdot10^{-22}$ and 4.92$\cdot10^{-23}$~cm$^2$ for [OI]~6300, [OI]~5577, [SII]~4068 and [SII]~6730, respectively), we assume a fixed dust-to-gas mass ratio of 10$^{-2}$ for the entire grid and a standard MRN type distribution \citep{Mathis1977}. Scattering is neglected. Recent work \citep{Franz2020} has shown that only grains smaller than approximately 10~$\mu$m can be entrained in a photoevaporative flow \citep[see also][]{Owen2011, Hutchison2016}. However we use the MRN distribution here as we are interested in an upper limit on the effect of dust absorption of these lines. As expected the effect is small.

Since our MHD model does not include a thick disc, we use the grids of the column number densities that we calculated for the photoevaporative model and remap them to be compatible with the MHD model.

To calculate the spectral profile we follow \citetalias{Ercolano2010} and \citetalias{Ercolano2016}. For each cell in our 3D-grid we calculate the line-of-sight velocity of the gas $v_{los}$, the thermal rms velocity of the emitting atom $v_{th}$ and the optical depth $\tau = C_{\lambda}^{abs} N_H$. We then create bins of velocity u with a resolution of 0.25 km/s and numerically integrate the luminosity in each bin by summing the contribution of the volume-averaged power $l(\vec{r})$ emitted in each cell of the 3D-grid:
\begin{align}
	L(u) = \int{d\vec{r} \frac{l(\vec{r})}{\sqrt{2\pi v_{th}(\vec{r})}}} exp\bigg( - \frac{[u - v_{los}]^2}{2 v_{th}(\vec{r})^2} -\tau (\vec{r}) \bigg),
\end{align}
where $\vec{r}$ is the vector pointing to the grid cell. We convolve our profiles with a Gaussian of width $\sigma = c / R$, where we choose the spectral resolution $R = 50000$, a resolution similar to the observations that we compare our profiles to. The observational sample is comprised of the high resolution ($\sim$7 km/s) observations analyzed by \citet{Fang2018} and \citetalias{Banzatti2018}. To facilitate the comparison we include observed profile components only if they are classified as either NLVC, BLVC or HVC.

We use a similar approach as \citetalias{Banzatti2018} to create fits to our synthetic profiles: We perform multi-Gaussian fits, beginning with a single Gaussian component and adding another component only if it improves the $\chi^2$ value by at least 20~\%, up to a maximum of 6 components. Since our profiles are much smoother and thus compare better to Gaussian components than real observed spectra, we introduce an additional exit condition: If $\chi^2 < \frac{2}{3}$~y$_{\mathrm{max}}^2$, where y$_{\mathrm{max}}$ is the flux at the peak of the profile, the fit is considered to be sufficiently accurate and no more components will be added. This is to prevent the algorithm from fitting to a too high level of detail that cannot be achieved in observations. Figures of all fits are available as online material. We adopt the categorization of \citetalias{Banzatti2018}, where all components with a blue- or redshift of more than 30~km/s are classified as HVC. Components slower than that are either classified as NLVC if their FWHM does not exceed 40~km/s or as BLVC, otherwise. We do not adopt the further distinction between BLVC+NLVC and SC or SCJ components, because correlations with the n$_{13-31}$ infrared-index suggest that single components could trace winds in more evolved discs, such as transition discs \citepalias{Banzatti2018}, whereas in this work we model primordial discs. We do note, however, that our photoevaporation model can only produce SC-type components, and that only the MHD model can produce more complex profiles with multiple components.

\section{Results}\label{sec:results}
On the basis of kinematic connections between the properties of BLVCs and NLVCs of the [OI]~6300 line, \citetalias{Banzatti2018} suggest both components to have their origin in the same MHD outflow, because a photoevaporative wind is launched at radii too large to reproduce the observed width of the broad components (assuming that these are Keplerian broadened). The existence of the observed connections had to be demonstrated for the scenario in which the NLVC originates in an outflow that is launched by a different mechanism, namely photoevaporation. The connections are in particular 
\begin{enumerate*}[label=(\arabic*)]
    \item a positive correlation between the quivalent widths of BLVCs and NLVCs.
    \item a positive correlation between the centroid velocities of BLVCs and NLVCs and 
    \item a positive correlation between the FWHMs of BLVCs and NLVCs
\end{enumerate*}

We can show with a simple correlation analysis (see appendix \ref{sec:appendix:correlations}) that these connections can be well explained by a single common correlation with a third variable, the accretion luminosity. We will show here that the photoevaporative wind model can reproduce the correlations with the accretion luminosity and disc inclination that are observed for the NLVCs and consequently that the observed connections between the [OI]~6300 BLVC and NLVC components can indeed be reproduced even when the NLVCs and BLVCs trace different wind types.

For both, the photoevaporative and the MHD wind models, we calculated the line emission for the three different accretion luminosities shown in table \ref{tab:input_parameters}. We calculated the resulting line profiles for four different lines, [OI]~6300, [OI]~5577, [SII]~4068 and [SII]~6730, each when viewed at an inclination of 0\degree{}, 20\degree{}, 40\degree{}, 60\degree{} and 80\degree{}. We will show maps of the wind structures and emission regions to provide insight into the physical conditions that are required for the emission of the different lines. We will use the results from the MHD model to demonstrate how even our very simple model can reproduce the observed components well and that the different components do not necessarily provide clear distinctions between physical regions of the outflow.

We repeated our analysis for a model that includes an additional illuminating EUV source in a fictional jet and we will show that such a "lamppost"-type illumination can induce line emission in higher regions of the photoevaporative wind, increasing the centroid velocity of the line profile.

\subsection{Emission regions} \label{sec:res:emission_regions}

\begin{figure*}
    \centering
    \includegraphics[width=\textwidth]{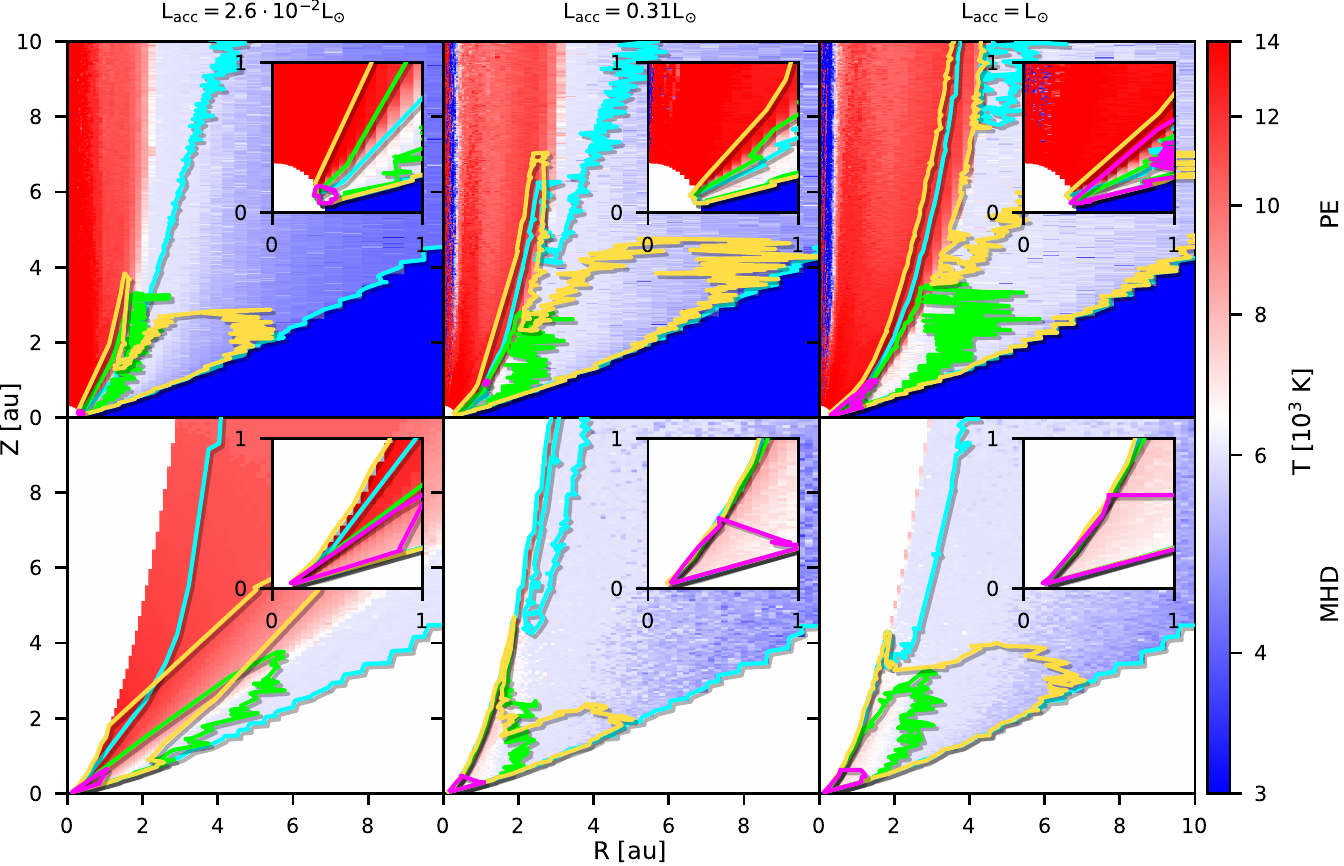}
    \caption{Temperature of the photoevaporative wind model (top panels) and the MHD wind model (bottom panels). Overlain are the contours of the 80\% emission regions of [OI]~6300 (green), [OI]~5577 (magenta), [SII]~4068 (yellow) and [SII]~6730 (cyan) for an accretion luminosity of $2.6\cdot 10^{-2} ~\mathrm{L_{\odot}}$ (left panels), $0.31~\mathrm{L_{\odot}}$ (middle panels) and 1~L$_{\odot}$ (right panels).}
    \label{fig:T_emission_all}
\end{figure*}

\begin{figure*}
    \centering
    \includegraphics[width=\textwidth]{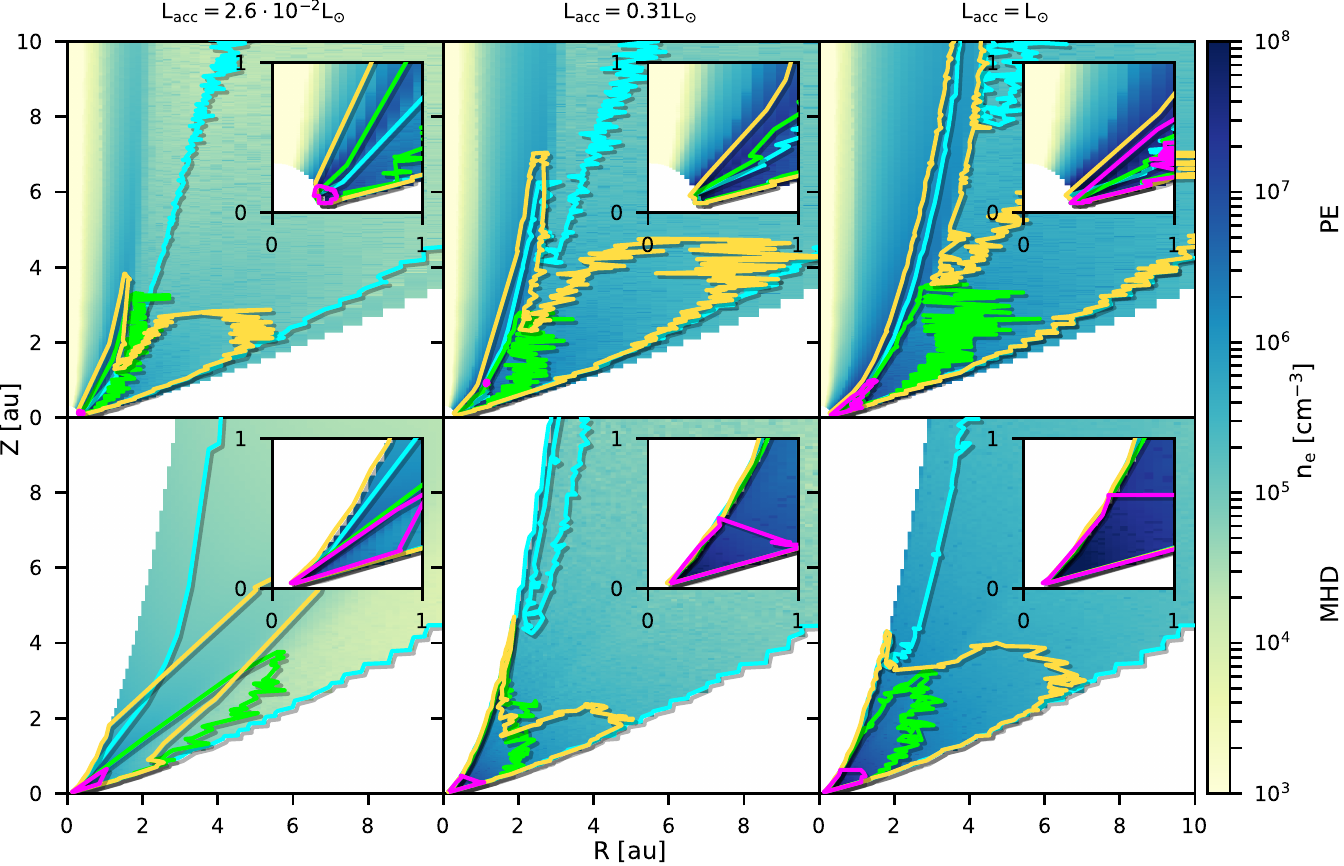}
    \caption{Electron number density of the photoevaporative wind model (top panels) and the MHD wind model (bottom panels). Overlain are the contours of the 80\% emission regions of [OI]~6300 (green), [OI]~5577 (magenta), [SII]~4068 (yellow) and [SII]~6730 (cyan) for an accretion luminosity of $2.6\cdot 10^{-2} ~\mathrm{L_{\odot}}$ (left panels), $0.31~\mathrm{L_{\odot}}$ (middle panels) and 1~L$_{\odot}$ (right panels).}
    \label{fig:ne_emission_all}
\end{figure*}

\begin{figure*}
    \centering
    \includegraphics[width=\textwidth]{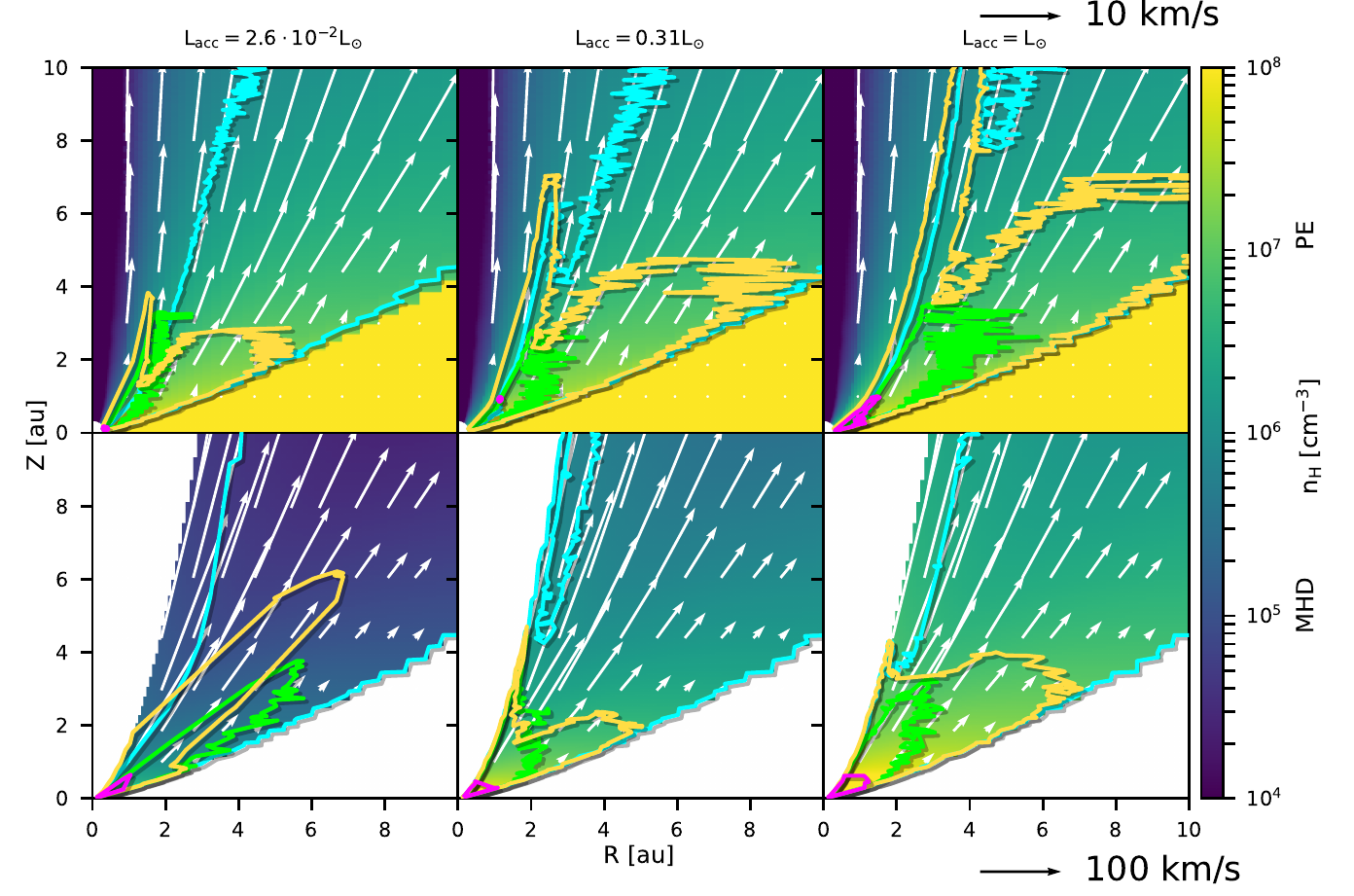}
    \caption{Density structure of the photoevaporative wind model (top panels) and the MHD wind model (bottom panels). The arrows show the velocity vectors in the RZ-plane. Overlain are the contours of the 80\% emission regions of [OI]~6300 (green), [OI]~5577 (magenta), [SII]~4068 (yellow) and [SII]~6730 (cyan) for an accretion luminosity of $2.6\cdot 10^{-2} ~\mathrm{L_{\odot}}$.}
    \label{fig:nh_v_emission_all}
\end{figure*}

\citet{Ercolano2016} have shown that their X-ray photoevaporative wind model can successfully reproduce the correlation between accretion luminosity and [OI]~6300 line luminosity by comparing their simulations with observations of \citet{Rigliaco2013} and \citet{Natta2014}. They suggest that this correlation is a consequence of the size of the wind region that the EUV flux, which is dominated by and proportional to the accretion luminosity, can reach and heat up. Since the luminosity of collisionally excited lines depends exponentially on the temperature through the Boltzmann term $\mathrm{exp(-\Delta E/k_B T)}$, a larger heated region leads to increased line luminosities. Figure \ref{fig:T_emission_all} shows temperature maps of the different models with the 80\% emission regions of all four lines overlain. Since the MHD model does not have a thick disc, we masked away the emission that would originate within the bound disc. The mask was constructed by remapping the bound disc of the photoevaporative model to the grid of the MHD model. It is clear from the figure that a higher accretion luminosity increases the size of the emission region. Another important factor for the luminosity of collisionally excited lines is the abundance of the emitting species (e.g. neutral oxygen for [OI] lines) and of the colliding particles that excite the species, which in our case are electrons and neutral hydrogen. It is important to note that our photoionisation calculations include neutral hydrogen as colliding particles only for the [OI]~6300 line, due to the lack of collisional rates with neutral hydrogen for the relevant excitation levels of the other lines. The calculated luminosities of the [OI]~5577, [SII]~4068 and [SII]~6730 should thus be regarded as a lower limit. The rate coefficients for [OI]~6300 are taken from \citet{Launay1977}, who report only the first four levels, $^3$P$_2$, $^3$P$_1$, $^3$P$_0$ and $^1$D$_2$, but not the $^1$S$_0$ level that is relevant for [OI]~5577. We verified that this only affects the total line luminosity but has no significant effect on the shape of the line profiles by recalculating the [OI]~6300 profiles with neutral hydrogen collisions turned off and comparing them to the original profiles. Figure \ref{fig:ne_emission_all} shows maps of the electron number density in our models and figure \ref{fig:nh_v_emission_all} shows maps of the hydrogen number density and the outflow velocities in the rz-plane. From these three figures we can read off the physical properties of the emission regions. The emission only becomes significant above a critical density, where the rate of collisional excitation to the upper level of the relevant transition matches the rate of radiative de-excitation. If the density is too low, not enough atoms/molecules are in the excited state. If it is too high, collisional de-excitation will dominate and suppress the transition. The temperature and the electron and hydrogen number densities in the emission regions are listed in table \ref{tab:emission_properties}. The [OI] 5577 line is only emitted in a small region close to the star where the density is comparatively high. As a result the line is very weak, which makes observations of this line challenging. The [SII]~6730 line is emitted even at very low densities and temperatures and therefore arises in a much larger region that reaches deep into the wind. Of our four lines it would be the best candidate to trace an extended disc wind.


\begin{table}
    \centering
    \begin{tabular}{cccc}
    Line & T [10$^3$K] & n$_\mathrm{e}$ [cm$^{-3}$] & n$_\mathrm{H}$ [cm$^{-3}$] \\
    \hline 
    \hline \relax
    [OI] 6300  & 6 -- 7     & 5$\cdot10^4$ -- 10$^8$    & 2$\cdot10^5$ -- 5$\cdot10^7$ \\\relax
    [OI] 5577  & 7 -- 8     & 5$\cdot10^6$ -- 10$^8$    & $10^8$ -- 10$^{10}$ \\\relax
    [SII] 4068 & 5 -- 12    & $10^5$ -- 10$^8$          & $10^5$ -- 5$\cdot10^7$ \\\relax
    [SII] 6730 & 4 -- 11    & $10^3$ -- 10$^8$          & $10^4$ -- 10$^7$ \\
\end{tabular}
    \caption{Temperature, electron number density and hydrogen number density of the line emission regions.}
    \label{tab:emission_properties}
\end{table}

\subsection{Line profiles} \label{sec:res:profiles}

\begin{figure*}
    \centering
    \includegraphics[width=\textwidth]{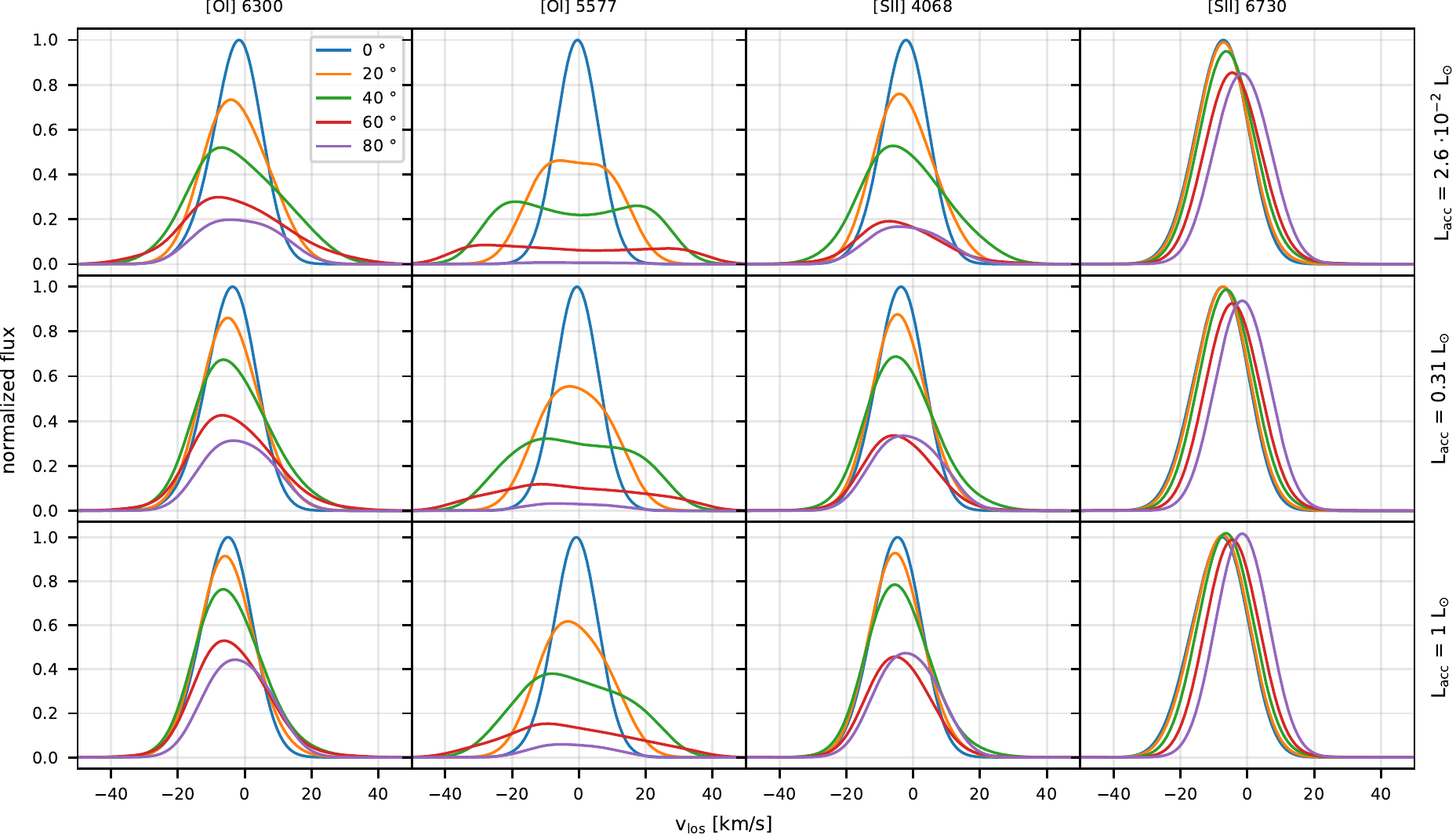}
    \caption{Simulated line profiles for the photoevaporative wind models. The profiles were artificially degraded to a spectral resolution of R = 50000 and normalized to the peak flux of the 0\degree{} profile.}
    \label{fig:profiles_pe}
\end{figure*}

The line profiles resulting from the X-ray photoevaporative wind models are shown in figure \ref{fig:profiles_pe}, normalized to the peak flux of the 0\degree{} inclination profiles. It is clear that the [OI]~6300 and [SII]~4068 profiles are very similar although the [SII]~4068 emission is reaching farther into the wind and along the inner edge of the wind. As can be seen in figure \ref{fig:nh_v_emission_all} (top panels) the regions have a very similar velocity structure, which explains the similarity of the line profiles. They both have blueshifted peaks of a few~km/s with the highest blueshift at 60\degree{} inclination. The [SII]~6730 profile is emitted up to higher regions and therefore less affected by dust attenuation at higher inclinations. The broadest profiles, when observed at an inclination, are those of the [OI]~5577 line. This is due to the very small emission region that lies very close to the star and is thus subject to strong Keplerian broadening. Although the other lines are emitted in that region, too, the contribution to their line profiles manifests only weakly in their wings. Unsurprisingly, the [OI]~5577 line is also the weakest and shows very little blueshift, because a photoevaporative wind cannot effectively be launched that close to the star. We fitted the profiles as described in section \ref{sec:met:profiles} and show the resulting component properties in table \ref{tab:fits_pe} in appendix \ref{sec:appendix:fits}. Figures of all fits are available as online material.

\begin{table*}
    \centering
    \begin{tabular}{cccccccc}
    & & \multicolumn{2}{c}{NLVC} & \multicolumn{2}{c}{BLVC} & \multicolumn{2}{c}{HVC} \\ [1.5ex]
    Model & Line 
        & v$_\mathrm{c}$ [km/s] & FWHM [km/s] 
        & v$_\mathrm{c}$ [km/s] & FWHM [km/s]
        & v$_\mathrm{c}$ [km/s] & FWHM [km/s] \\
    \hline
    \hline
    \multirow{4}{*}{\rotatebox[origin=c]{90}{Photoevap.}}
        & [OI] 6300  & -3.74 $\pm$  1.22 & 24.30 $\pm$ 5.25 \\
        & [OI] 5577  &  0.78 $\pm$ 11.49 & 25.65 $\pm$ 6.60 & -2.84 $\pm$ 0.91 & 43.56 $\pm$ 2.66 & -30.11 $\pm$ 0 & 19.92 $\pm$ 0 \\
        & [SII] 4068 & -3.71 $\pm$  1.14 & 22.13 $\pm$ 3.85 \\
        & [SII] 6730 & -5.51 $\pm$  2.34 & 19.03 $\pm$ 0.41 \\
    \hline
    \multirow{4}{*}{\rotatebox[origin=c]{90}{MHD}} 
        & [OI] 6300  & -2.09 $\pm$ 17.11 & 26.15 $\pm$ 5.40 & -14.30 $\pm$ 21.87 & 58.56 $\pm$  9.78 & -100.05 $\pm$ 79.90 &  92.78 $\pm$ 53.62 \\
        & [OI] 5577  & -7.70 $\pm$ 14.10 & 28.13 $\pm$ 4.08 &  28.10 $\pm$ 0.91  & 47.54 $\pm$  3.70 &  -60.81 $\pm$ 49.19 &  93.23 $\pm$ 43.81 \\
        & [SII] 4068 & -3.66 $\pm$ 16.16 & 23.81 $\pm$ 3.65 & -10.94 $\pm$ 18.79 & 52.97 $\pm$ 15.28 &  -89.93 $\pm$ 44.01 & 123.37 $\pm$ 59.24 \\
        & [SII] 6730 & -3.13 $\pm$ 12.86 & 22.78 $\pm$ 5.35 & -17.70 $\pm$ 12.31 & 55.91 $\pm$ 22.53 &  -92.57 $\pm$ 52.24 & 113.71 $\pm$ 57.91 \\
\end{tabular}
    \caption{Mean values of the centroid velocities and FWHM and their standard deviation for all lines and components.}
    \label{tab:component_stats}
\end{table*}

\begin{figure*}
    \centering
    \includegraphics[width=\textwidth]{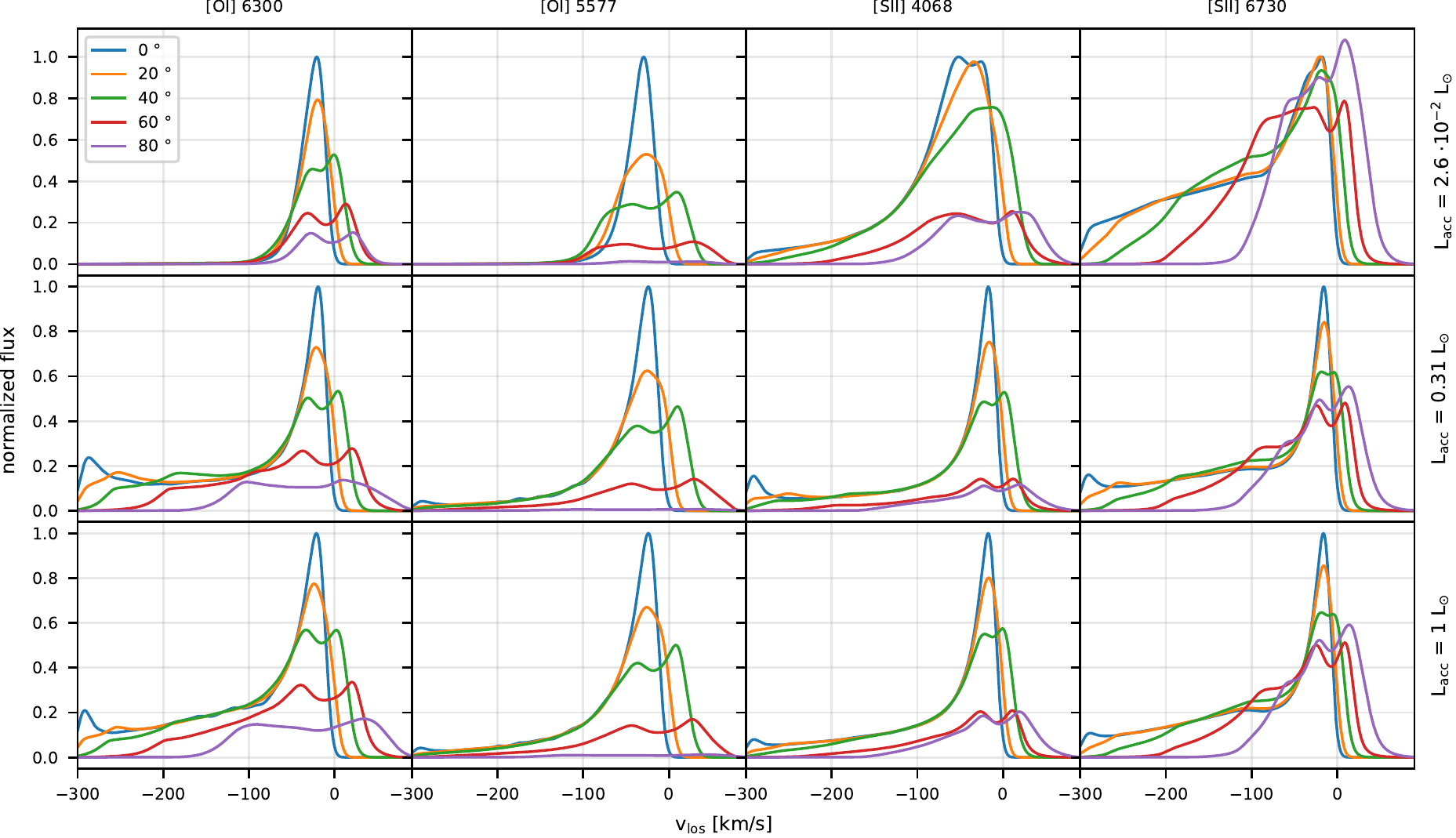}
    \caption{Simulated line profiles for the MHD wind models. The profiles were artificially degraded to a spectral resolution of R = 50000 and normalized to the peak flux of the 0\degree{} profile.}
    \label{fig:profiles_mhd}
\end{figure*}

\begin{figure}
    \centering
    \includegraphics[width=.48\textwidth]{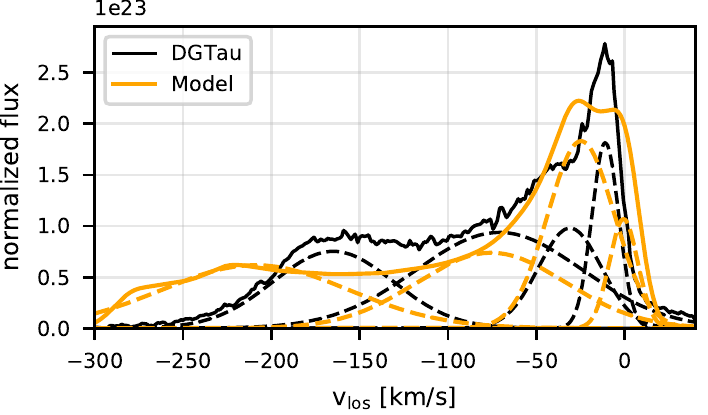}
    \caption{Simulated [OI]~6300 profiles (orange lines) of the MHD wind model with an accretion luminosity of $0.31~\mathrm{L_{\odot}}$ at 30\degree{} inclination compared to the observed spectrum of DG Tau (black lines). The dashed lines show the individual components from the Gaussian decomposition. The synthetic profile has been artificially degraded to a spectral resolution of R = 50000. The profiles have been normalized in such a way that they are comparable to each other. The DG Tau spectrum is taken from \citetalias{Banzatti2018}.}
    \label{fig:DGTau_comp_mhd}
\end{figure}

Figure \ref{fig:profiles_mhd} shows the profiles for the MHD wind model. By far the most notable difference when comparing these profiles to the photoevaporative wind model are the very high blueshifts. At higher accretion luminosities and mass-loss rates all profiles show significant blueshifts up to velocities of $\sim$300~km/s. Despite the simple model, some of the profiles are remarkably similar to observed profiles, although the model generally underestimates the line luminosities. Figure \ref{fig:DGTau_comp_mhd} compares the 30\degree{} inclined [OI]~6300 profile to the profile of DG Tau, which has a similar inclination. Note that using the [OI]~6300 luminosity of DGTau from \citet{Simon2016a}, we find that the luminosity of our synthetic line is a factor of $\sim$25 too low. The profiles have thus been normalized to account for this factor. We highlight that our work shows that the observed HVC can be reproduced by our analytical MHD wind without the need of explicitly including a 'classical' collimated jet. Indeed, with blueshifts up to $\sim$300~km/s, the high-velocity wings extend to even higher blueshifts than the wing of the DG Tau profile. In the understanding that the fastest, innermost edge of the wind model is a jet, this demonstrates that the jet does not need to be collimated or exhibit shocks in order to produce the observed HVCs, as proposed in previous works. At lower blueshifts the emission of our synthetic profile is too broad and with hints of Keplerian double peaks compared to that of DG Tau. In fact, we find pronounced double peaks in all our inclined profiles, but they are rarely observed. This could indicate that the Keplerian broadening is too strong in our model, but this is unlikely considering the inner radius of 20~R$_{\odot}$, which is relatively large for an MHD wind. An alternative explanation is that the Keplerian throught is filled out by another narrow low velocity component originating in a different outflow, such as a photoevaporative wind. We explore this possibility in section \ref{sec:discussion:profiles_combi}. The reduced high-velocity flux in the profiles of the model with the lowest accretion rate can be easily understood with the help of the emission maps in figures \ref{fig:T_emission_all} -- \ref{fig:nh_v_emission_all}: The density in the fast flowing inner region of the inner wind is low. As a result the temperature is high, as is the degree of ionization. With the low density and high degree of ionization, not enough neutral oxygen is present for the [OI] lines to be emitted. The singly ionized sulfur is still abundant enough to produce a significant flux of the [SII] lines in the high-velocity region. As the accretion luminosity and with it the density of the wind increases, the fast moving inner edge is less ionized and the high-velocity flux increases. The [OI]~5577 line that is only emitted close to the star has no significant contribution in the high-velocity region. The MHD profiles are typically best fitted with 3 to 5 components. All fitted components are listed in tables \ref{tab:fits_mhd_pt1} and \ref{tab:fits_mhd_pt2}.

Table \ref{tab:component_stats} shows the mean values of the centroid velocity and FWHM of the NLVCs and where applicable the BLVCs and HVCs for the photoevaporative and MHD models. Table \ref{tab:luminosities} shows the line luminosities.

\begin{table*}
    \centering
    \begin{tabular}{ccccccccc}
    \multicolumn{1}{l}{Model} & & \multicolumn{3}{c}{Photoevaporation} & & \multicolumn{3}{c}{MHD} \\ [1.5ex]
    \multicolumn{1}{l}{L$_\mathrm{acc}$ [L$_{\odot}$]} 
        & & $2.6\cdot10^{-2}$ & 0.31 & 1
        & & $2.6\cdot10^{-2}$ & 0.31 & 1 \\
    \hline
    \hline
    \multirow{4}{*}{\rotatebox[origin=c]{90}{[OI] 6300}}
        & 0 \degree  & -4.82 & -4.32 & -3.90 & & -5.54 & -4.54 & -3.87 \\
        & 20 \degree  & -4.82 & -4.32 & -3.90 & & -5.54 & -4.55 & -3.88 \\
        & 40 \degree  & -4.82 & -4.32 & -3.90 & & -5.55 & -4.55 & -3.88 \\
        & 60 \degree  & -5.03 & -4.49 & -4.03 & & -5.74 & -4.74 & -4.07 \\
        & 80 \degree  & -5.28 & -4.66 & -4.13 & & -6.00 & -5.03 & -4.32 \\
    \hline
    \multirow{4}{*}{\rotatebox[origin=c]{90}{[OI] 5577}}
        & 0 \degree  & -5.87 & -5.36 & -4.94 & & -7.07 & -5.62 & -4.75 \\
        & 20 \degree  & -5.88 & -5.37 & -4.95 & & -7.09 & -5.63 & -4.76 \\
        & 40 \degree  & -5.89 & -5.38 & -4.96 & & -7.10 & -5.64 & -4.77 \\
        & 60 \degree  & -6.31 & -5.76 & -5.31 & & -7.49 & -6.03 & -5.14 \\
        & 80 \degree  & -7.62 & -6.57 & -5.94 & & -8.46 & -7.23 & -6.17 \\
    \hline
    \multirow{4}{*}{\rotatebox[origin=c]{90}{[SII] 4068}}
        & 0 \degree  & -4.48 & -3.85 & -3.47 & & -5.39 & -4.34 & -3.55 \\
        & 20 \degree  & -4.48 & -3.85 & -3.47 & & -5.39 & -4.35 & -3.55 \\
        & 40 \degree  & -4.49 & -3.86 & -3.47 & & -5.40 & -4.36 & -3.56 \\
        & 60 \degree  & -4.97 & -4.17 & -3.71 & & -5.83 & -4.84 & -3.94 \\
        & 80 \degree  & -5.04 & -4.16 & -3.68 & & -5.89 & -4.95 & -3.97 \\
    \hline
    \multirow{4}{*}{\rotatebox[origin=c]{90}{[SII] 6730}}
        & 0 \degree  & -4.70 & -4.33 & -4.05 & & -5.95 & -5.18 & -4.55 \\
        & 20 \degree  & -4.70 & -4.32 & -4.05 & & -5.95 & -5.18 & -4.55 \\
        & 40 \degree  & -4.70 & -4.32 & -4.05 & & -5.95 & -5.18 & -4.55 \\
        & 60 \degree  & -4.74 & -4.35 & -4.07 & & -5.99 & -5.25 & -4.61 \\
        & 80 \degree  & -4.74 & -4.35 & -4.07 & & -6.01 & -5.28 & -4.64 \\
\end{tabular}
    \caption{Logarithm of the line luminosities in units of~L$_{\odot}$.}
    \label{tab:luminosities}
\end{table*}

\subsubsection{Line profile decomposition} \label{sec:res:profile_decomposition}

\begin{figure*}
    \centering
    \includegraphics[width=.9\textwidth]{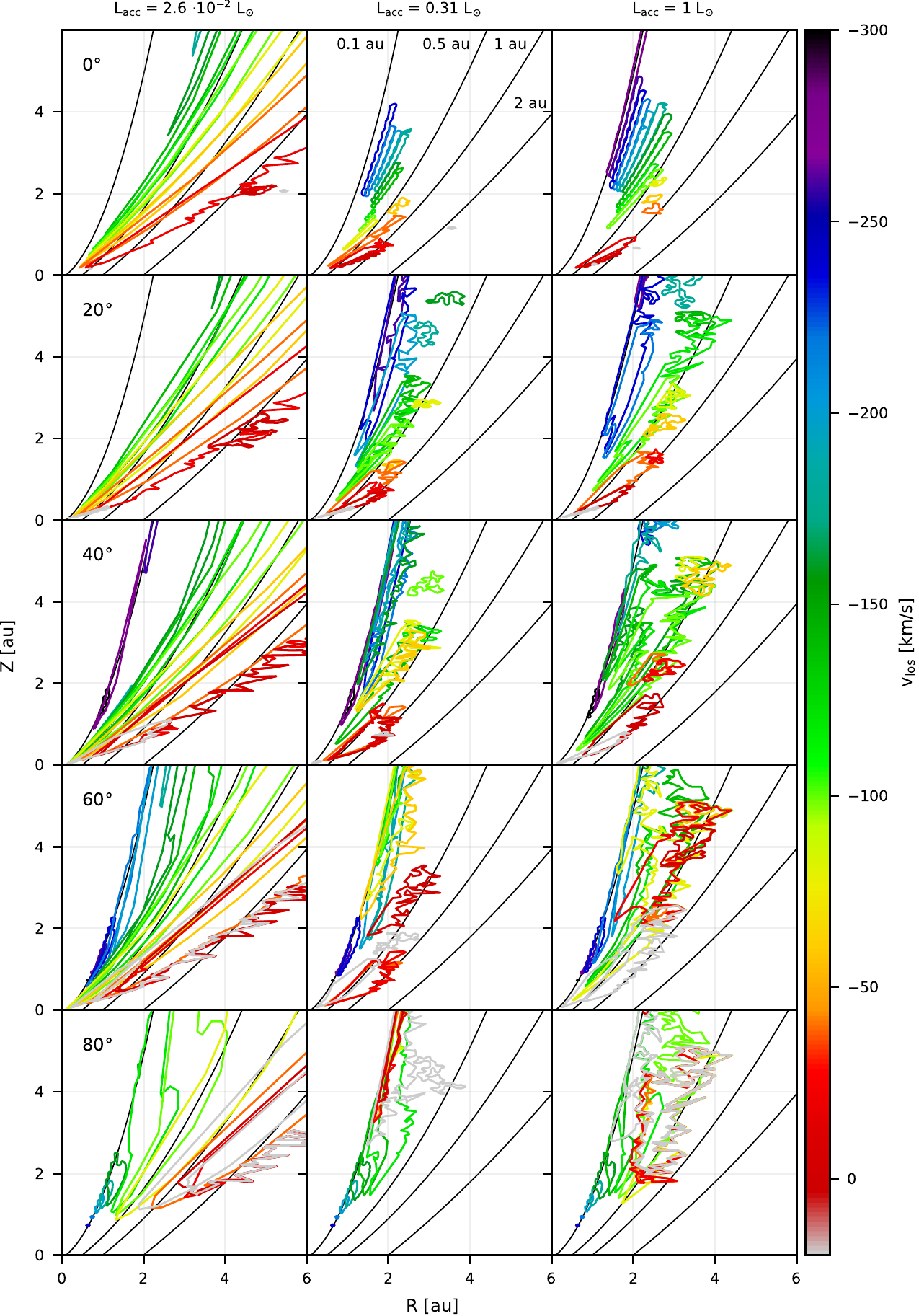}
    \caption{80~\% emission regions of [OI] 6300 flux that is observed at different velocities for the three MHD models with a spectral resolution of R = 50000.}
    \label{fig:OI-6300_v_emis}
\end{figure*}

It is true for all lines that the line broadening is entirely dominated by the velocity gradient of the wind, when the disc is viewed at low inclinations. In that case the profiles do not allow for a clear distinction between low-velocity flux originating close to the star in the slow base of a fast wind or jet and emission originating in a slow extended disc wind far away from the star. The formation of Keplerian double peaks at inclinations $\gtrsim$ 40\degree{} indicates that Keplerian rotation begins to affect the broadening of the low velocity-flux around that inclination. However, on the blueshifted side, an increase of the inclination will result in contamination of the low-velocity component with emission from high-velocity regions that is projected to the low-velocity regime. This contamination is strengthened by the fact that the high-velocity wind regions have the smallest outflow-angles with respect to the z-axis in the rz-plane, resulting in stronger projection effects for HVCs and stronger broadening for LVCs. To illustrate this, figure \ref{fig:OI-6300_v_emis} shows the 80~\% emission regions of the [OI]~6300 line in the MHD models for different velocities in increments of 20 km/s. At 0\degree{} inclination the regions are well separated and could be traced using the centroid velocity of narrow components, if they are no broader than $\sim$20~km/s. At higher inclinations, starting already at 20\degree{}, the emission regions from flux observed in the range between +40~km/s and -40~km/s overlap significantly. At high inclinations projection effects are strong and a distinction of emission regions based on the blue- or redshift is difficult. As a result, a fitted Gaussian component will contain flux from many different parts of the wind. We therefore advise against interpreting different fit components as tracing physically distinct outflow regions. For the wind models with higher accretion luminosities, it is true that the majority of the emission traces those parts of the wind that are launched inside 1 au or even 0.5 au at higher inclinations. In our fits these models do reproduce NLVCs that are consistent with observations and typically attributed to being launched at larger radii (c.f. figure \ref{fig:corr_inc}). This demonstrates that although the velocity slices do reflect the onion-like velocity structure of the wind, the line width of the components is not a good indicator for the emission region, neither when the disc is viewed at low inclinations, nor at high inclinations. Line ratios between components should only be considered with caution, because the components are likely to contain flux from multiple regions with different physical properties. We expect the centroid velocity, as a proxy for the velocity slice, to hold most of the physical meaning of the Gaussian components. It could be worth considering alternative measurements (e.g. a spectral slope) for future studies of correlations between line profiles and physical properties of the disc.

\begin{figure*}
    \centering
    \includegraphics[width=.9\textwidth]{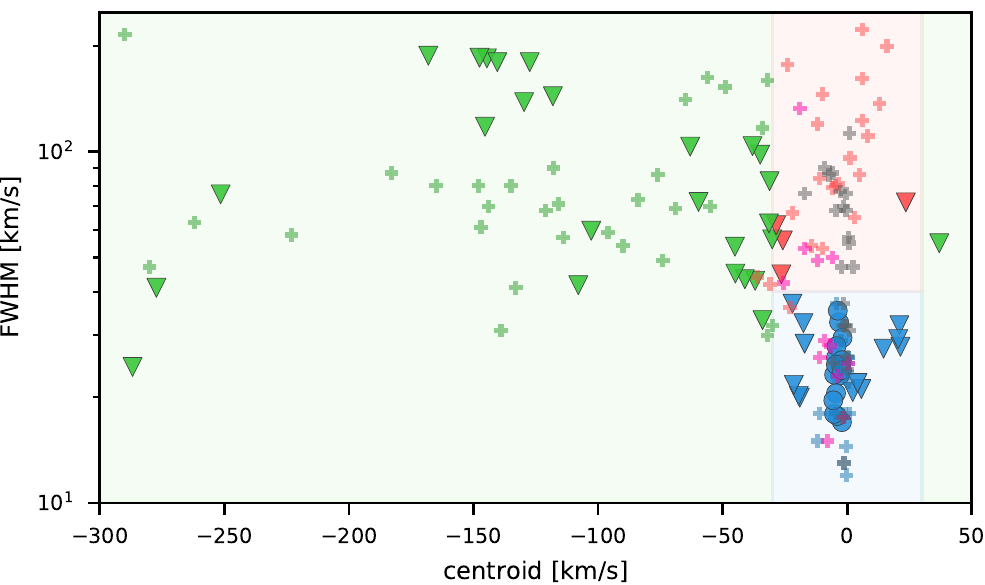}
    \caption{Overview of the parameter space of centroid velocities and FWHM of the [OI]~6300 profile components. Components from the MHD models are marked with a triangle, those from the photoevaporation model with a circle. The plusses mark observations by \citetalias{Banzatti2018}. The colors indicate the type of the components: Green: HVC, red: BLVC, blue: NLVC, grey: SC, pink: SCJ.}
    \label{fig:OI-6300_overview}
\end{figure*}

Nevertheless the Gaussian fits to observed profiles do reveal some correlations. We can compare our fits with the observations, in order to see if we can reproduce the general trends. Figure \ref{fig:OI-6300_overview} shows an overview of the centroid velocities and FWHM of the [OI]~6300 line.  Except for the narrowest observations and the two observations with a blueshift around $\sim$12~km/s, the NLVCs are well reproduced by the photoevaporation models. The SCs, while having similar blueshifts, are typically much broader than the NLVCs and cannot be reproduced by our photoevaporation model. Correlations with infrared-index reported by \citetalias{Banzatti2018} suggest that these components could trace winds in more evolved discs, such as transition discs. As has been shown by \citetalias{Ercolano2016}, transition disc profiles are generally broader, although still not broad enough to match the observed SC line widths. 
BLVCs and HVCs are only produced by the MHD model. The HVCs tend to be too broad around a blueshift of $\sim$150~km/s when compared to observations. This discrepancy can be explained by the fitting procedure which prefers one single broad Gaussian over two narrower Gaussians that would better match the observations. The one redshifted HVC comes from a profile at 80\degree{} and is a fit to the redshifted part of the Keplerian double peak. The MHD model consistently overestimates the blue or redshift of the NLVCs. With only 4 BLVCs at the narrower end of the spectrum, the model clearly tends to favour NLVCs and HVCs over BLVCs. These discrepancies are a consequence of the Keplerian double peaks that are found in the majority of our profiles. They are better fitted by two narrow components than a broad component.

\subsection{Correlations with inclination} \label{sec:res:correlations_inc}

\begin{figure*}
    \centering
    \includegraphics[width=.8\textwidth]{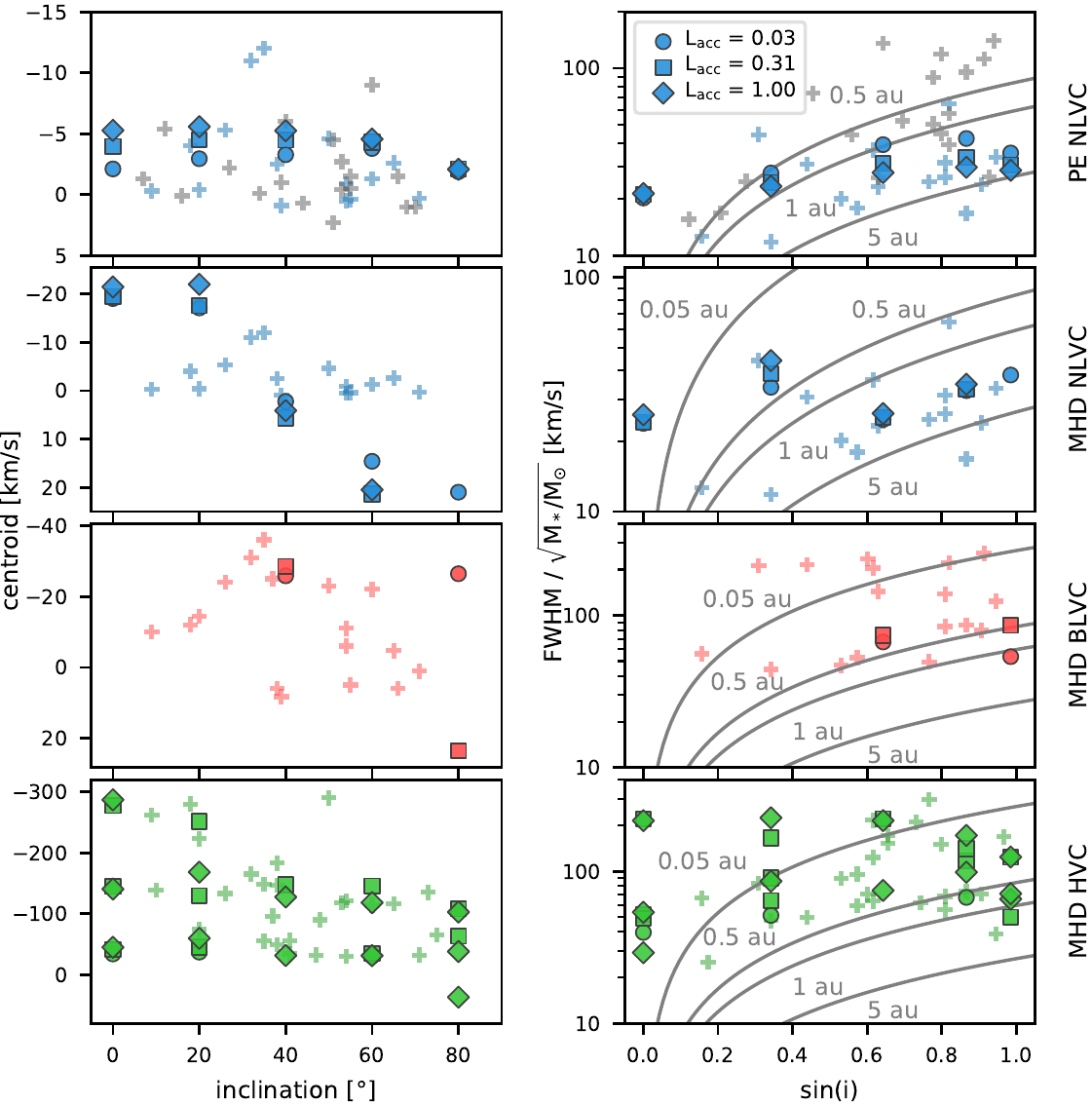}
    \caption{Correlations between the disc inclination and the [OI]~6300 line component centroid (left panels) and FWHM normalized with the square root of the stellar mass (right panel). The grey lines show the values expected from purely Keplerian broadening at different radii. The plusses mark observations by \citetalias{Banzatti2018}. The colors indicate the type of the components: Green: HVC, red: BLVC, blue: NLVC, grey: SC.}
    \label{fig:corr_inc}
\end{figure*}

\citetalias{Banzatti2018} investigate the correlation between [OI]~6300 centroid velocities and viewing angle in their sample and find a weak trend of the LVCs to have their maximum centroid velocity at an inclination around 35\degree{}. As can be seen in figure \ref{fig:corr_inc} none of our models is able to clearly reproduce this trend. The top left panel shows the inclination against centroid velocities of our photoevaporation model profile components compared to the observed NLVCs. While the model with the lowest accretion luminosity has its peak at 60\degree{} inclination, the models with higher accretion luminosities show only a weak trend with a maximum around 20\degree{}. As is clear from the figure and as has already been pointed out by \citetalias{Banzatti2018} the centroid velocities are dominantly influenced by the accretion luminosity and not by the viewing angle. We will investigate the correlations between the component properties and accretion luminosity in the next section. We have seen in the previous section that the photoevaporation model fails to reproduce the two observed components with the highest blueshifts. As we will show in section \ref{sec:res:jet}, a lamppost-type illumination of the upper layers of the X-ray photoevaporative wind by a UV source in a jet could help close the gap to these observations. The right panel shows the correlation of the component width with inclination. The widths are consistent with Keplerian broadening between 0.5 and 5~au, which is also consistent with the emission maps, but as seen at inclinations < 20\degree{}, Kepler broadening is not the dominating mechanism. 

The MHD model seems to produce NLVCs with decreasing blueshift that transitions into a redshift with increasing inclination. A visual inspection of the fits in appendix \ref{sec:appendix:fits} reveals that most profiles are fitted with a NLVC at their right edge, which explains this behaviour. The model is able to produce NLVCs that reach up to 30~km/s but similar to the photoevaporation model it cannot reproduce the narrowest of the observed NLVCS. At inclinations < 25\degree{} the observations show a reduced number of NLVCs and BLVCs with intermediate blueshifts of $\sim$10 - 30~km/s.  One possible cause could be the 30~km/s threshold above which components get classified as HVCs. Outflows in a jet along the z-axis with velocities slightly above that threshold would have a projected velocity just below it, resulting in a classification as BLVC. If that was the case we would expect a reduced number of HVCs at higher viewing angles but this is difficult to assess with the given distribution of observational samples. In fact, it is much more likely that the low number of components at low inclinations is a direct consequence of the non-uniform distribution of samples, as is shown in figure \ref{fig:inclinations}. This could also explain the observed peak at $\sim$35\degree{}. The general lack of BLVCs makes it impossible to obtain useful information about the effect of the inclination on the BLVCs. The centroid velocities of the HVCs generally decrease with increasing inclination, matching the observations well. The figure helps to support two arguments that were made in the previous section: The redshifted HVC belongs to the 80\degree{} profile of the 1~L$_{\mathrm{acc}}$ model and the very broad HVCs are mostly present at low inclinations up to 40\degree{}. At higher inclination the extended blueshifted wing is projected to lower velocities and the HVCs become somewhat narrower.

\subsection{Correlations with accretion luminosity} \label{sec:res:correlations_acc}

\begin{figure*}
    \centering
    \includegraphics[width=\textwidth]{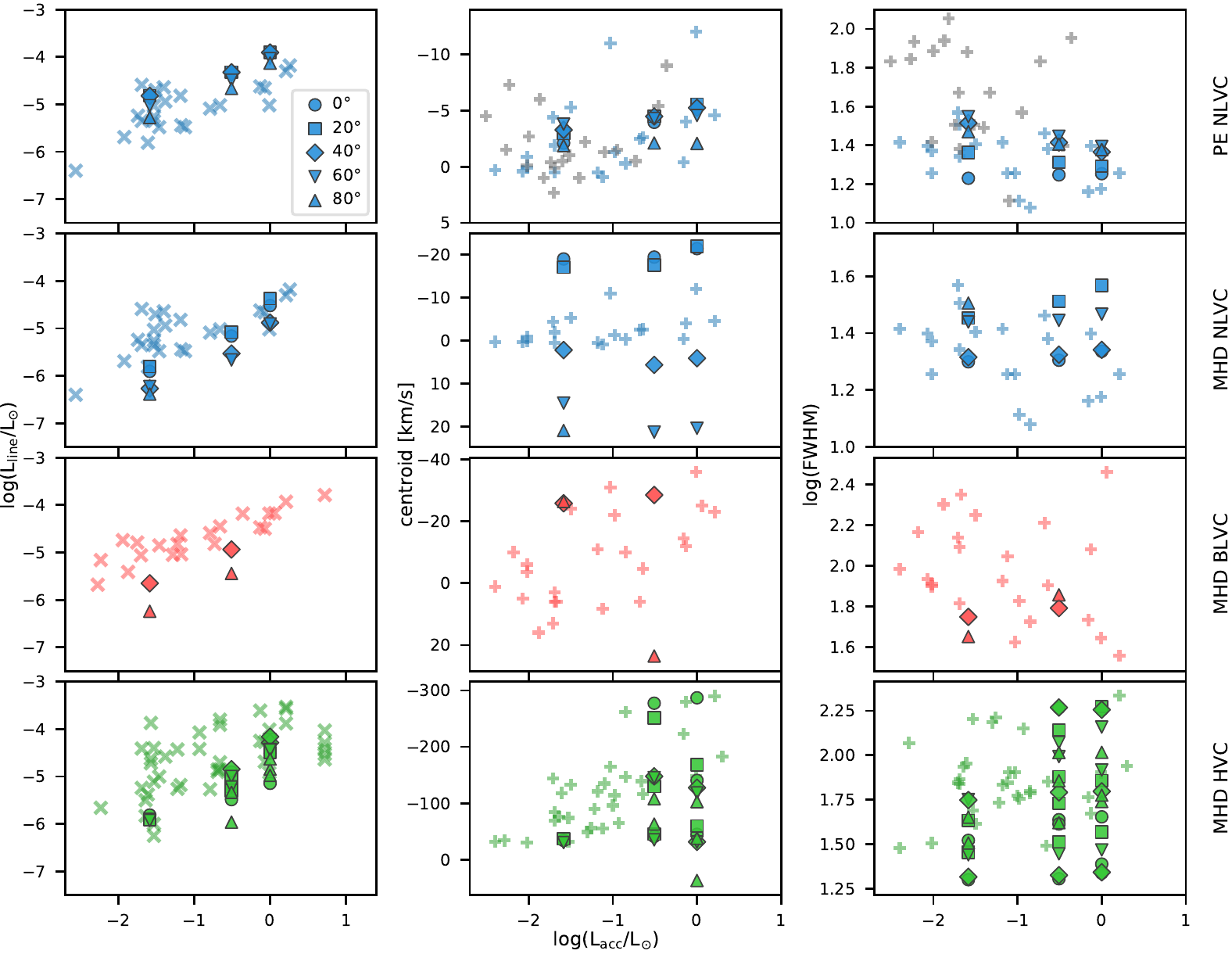}
    \caption{Correlations between the accretion luminosity and [OI]~6300 line component luminosities (left panels), centroids (middle panels) and FWHM (right panels). The plusses mark observations by \citetalias{Banzatti2018}, the crosses mark observations by Fang et al. (2018). The colours have the same meaning as in figure \ref{fig:corr_inc}. }
    \label{fig:corr_acc}
\end{figure*}

The more interesting correlations, which we suggest to be the underlying cause of all three NLVC-BLVC connections reported by \citetalias{Banzatti2018} are those between the component properties and the accretion luminosity. We have shown in appendix \ref{sec:appendix:correlations} that in order to reproduce these observed connections it is sufficient to reproduce the correlations with L$_{\mathrm{acc}}$. Figure \ref{fig:corr_acc} shows these correlations for our models and the comparison to the observations. The first reported connection is a positive correlation between the equivalent width of the NLVCs and that of the BLVCs. The NLVCs produced by our photoevaporative wind model are in good agreement with the observations, which shows that in a scenario where the NLVCs are produced by a photoevaporative wind and the BLVCs by a different wind type, one would still find the observed connection between the NLVC and BLVC luminosities. It is worth noticing at this point that the luminosities of our MHD model components are at or below the lower limit of the observed luminosities but the slope matches the observations. Compared to the photoevaporative wind model the densities in our MHD model are much lower for a wind with the same mass loss-rate. As a result the collisionally excited lines are weaker, too.

The second connection reported by \citetalias{Banzatti2018} is a positive correlation between the centroid velocity of the NLVCs and that of the BLVCs. As is shown in the middle panel of figure \ref{fig:corr_acc}, the NLVC centroids of our photoevaporative wind model increase with increasing accretion luminosities. This can be explained by the larger emission regions for the models with higher L$_{\mathrm{acc}}$, as we will discuss in section \ref{sec:discussion:correlations}. The correlation is again consistent with the observed NLVCs, which shows that the observed connection between NLVC and BLVC centroids could also be achieved when the two components trace different wind types. We have seen before that the MHD model does not reproduce the NLVC and BLVC centroids very well, but the HVC centroids tend to be more blueshifted with increasing accretion luminosity, although there are a number of HVCs with lower blueshifts than would be expected from the observations. It is likely that these components are another consequence of our fits to Keplerian double peaks.

Finally, \citetalias{Banzatti2018} reported a positive correlation between the FWHM of the NLVCs and that of the BLVCs. The observations show a decreasing FWHM with L$_{\mathrm{acc}}$ for both, the NLVC and the BLVC. The right panel of figure \ref{fig:corr_acc} clearly demonstrates that the FWHM of the NLVCs in our photoevaporative model matches this correlation. There is no clear trend in the observed HVCs but the MHD model does not have any broad HVCs in the model with low L$_{\mathrm{acc}}$ as a direct consequence of the lack of highly blushifted flux.

By showing that the photoevaporation model can reproduce the correlations of the NLVCs with L$_{\mathrm{acc}}$, we have demonstrated that all three reported NLVC-BLVC connections are compatible with a scenario in which the NLVC traces a photoevaporative wind and the BLVCs are produced in a different wind. The correlation between [OI]~6300 luminosity and L$_{\mathrm{acc}}$ does not distinguish between different wind scenarios, because both, photoevaporative and MHD winds produce such a positive correlation. This argues against conclusions \citep[e.g. by][]{Nisini2017} that this correlation alone suggests an MHD origin for the wind traced by [OI]~6300. What distinguishes different winds is truly the velocity structure and not necessarily the luminosity.

\subsection{Lamppost illumination from a jet}\label{sec:res:jet}

\begin{figure*}
    \centering
    \includegraphics[width=\textwidth]{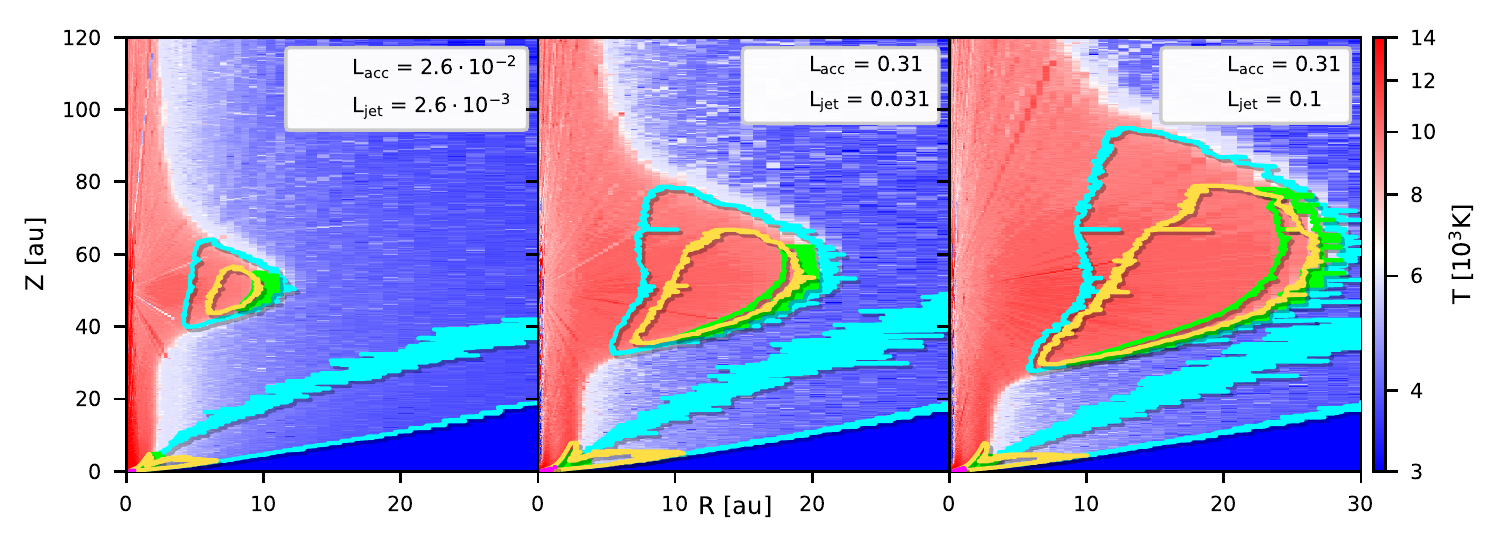}
    \caption{Electron number density (top) and temperature (bottom) of the photoevaporative wind model with accretion luminosity $2.6\cdot 10^{-2}~\mathrm{L_{\odot}}$ and an additional EUV jet source at Z = 50 AU on the z-axis that is modelled as a blackbody with T = 25000 K and luminosity $2.6\cdot 10^{-3}~\mathrm{L_{\odot}}$. Overlain are the contours of the 80\% emission regions of [OI]~6300 (green), [OI]~5577 (magenta), [SII]~4068 (yellow) and [SII]~6730 (blue) for an accretion luminosity of $2.6\cdot 10^{-2}~\mathrm{L_{\odot}}$ (left panels), $0.31~\mathrm{L_{\odot}}$ (middle panels) and $1~\mathrm{L_{\odot}}$ (right panels). The [OI]~5577 emission region lies entirely inside 1 au and is not visible at the scale of the figure.}
    \label{fig:T_emission_jet_pe}
\end{figure*}

\begin{figure}
    \centering
    \includegraphics[width=.48\textwidth]{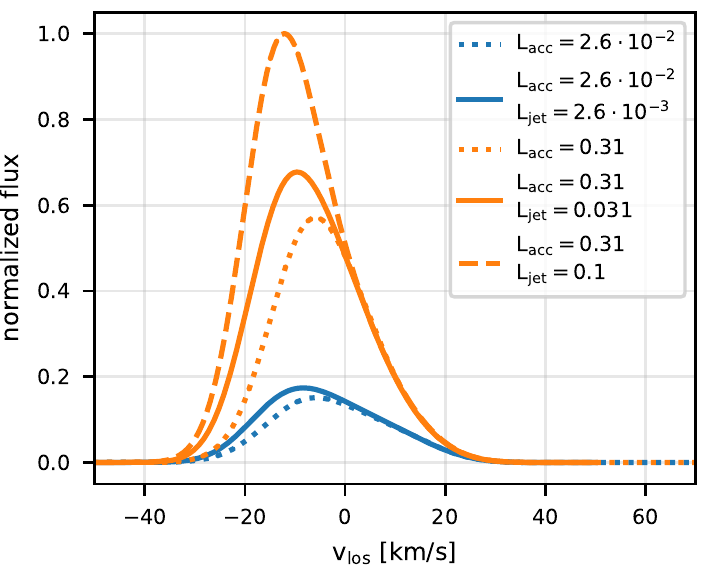}
    \caption{Simulated [OI]~6300 profiles at 30\degree{} inclination from the photoevaporative wind models with and without jet components. The dotted lines show the profiles without a jet component, the solid lines with a jet source that is 10~\% of the accretion luminosity. The luminosities in the legend are given in units of L$_{\odot}$. All jet components are modelled as a blackbody with T = 25000 K at a height of 50 AU on the z-axis. All profiles have been artificially degraded to a spectral resolution of R = 50000.}
    \label{fig:profiles_jet_pe}
\end{figure}

The observations show that a HVC is often present when both a broad and a narrow component is detected in the LVC of [OI] 6300. If the HVC is indicative of a jet, as often assumed in the literature, then it is useful to investigate how the profiles change when the wind is heated at higher heights above the midplane. To this aim, we modelled a scenario in which the wind is illuminated in a lamppost-type fashion from a source placed higher up on the z-axis. This source could potentially be an EUV or X-ray source created by shocks in a jet at a height of a few tens of au. We repeated our calculations for the photoevaporative and MHD models with an accretion luminosity of $2.6\cdot 10^{-2}~\mathrm{L_{\odot}}$ and 0.31~L$_\mathrm{{\odot}}$, both with an additional EUV input spectrum at a height of 50 au on the z-axis, which we modelled as a blackbody of temperature 25000 K and varying luminosity. As is shown in figure \ref{fig:T_emission_jet_pe}, upper wind levels can contribute significantly to the emission when the photoevaporative wind model is heated by an illumination source in the jet. Table \ref{tab:luminosities_pe_jet} shows the total line luminosities for the different lamppost-models and their ratio to their luminosity in the models without a jet. The [OI]~5577 line is not significantly affected by the jet source, as it requires a higher density to be emitted efficiently. The resulting [OI]~6300 profiles are shown in figure \ref{fig:profiles_jet_pe}. Compared to the profiles without a lamppost-type illumination, the centroid velocity of the fitted NLVC can increase the blueshift from $\sim$5~km/s to up to 10~km/s. The individual fits for the [OI]~6300 profiles are listed in table \ref{tab:fits_pe_jet}. With increasing luminosity of the jet source or decreasing accretion luminosity, the emission from the higher wind layers can dominate over the emission from lower layers, which will lead to more blueshifted but narrower profiles. This scenario is able to close the gap to the observed NLVCs with the highest blueshifts of up to $\sim$12~km/s. While the centroid velocities increase, the FWHM tend to decrese with increasing luminosity of the jet source. Apart from the [SII]~6730 line, the lamppost illumination has no effect on the emission in the MHD wind model. In that model, the density of the upper wind is too low for the other lines to be emitted efficiently. Since the MHD wind is very fast at high heights, the additional [SII]~6730 flux manifests in an increase in high-velocity flux. Because the velocity structure is relatively constant that high in the upper wind, the height of the jet source has little effect, as long as it is not placed low enough for the launching region of the wind to be heated by the jet source. The temperature from the jet source, however, does play an important role. A higher temperature implies a higher EUV flux, which is able to heat a larger volume of the wind, but also increases the degree of ionization, suppressing emission of neutral or lowly ionized species. X-ray sources, such as those observed in multiple jets \citep[e.g.][]{Gudel2007a} have little effect on the line profiles, as X-rays are mainly absorbed by heavier elements and not by hydrogen, making them inefficient in heating the wind.

\begin{table}
    \centering
    \begin{tabular}{cccccc}
    L$_{\mathrm{acc}}$ & L$_{\mathrm{jet}}$ & \multirow{2}{*}{Line} 
        & \multirow{2}{*}{log\big($\mathrm{\tfrac{L_{line}}{L_{\odot}}}$\big)}
        & \multirow{2}{*}{$\mathrm{\tfrac{L_{with ~jet}}{L_{without ~jet}}}$} \\\relax
    [L$_{\odot}$] & [L$_{\odot}$] & & & \\
    \hline
    \hline
    \multirow{4}{*}{$2.6\cdot10^{-2}$}
    & \multirow{4}{*}{$2.6\cdot10^{-3}$}
    	& \multirow{1}{*}{[OI]~6300}
    		& -4.75 & 1.17 \\
    	\cline{3-5}
    	& & \multirow{1}{*}{[OI]~5577}
    		& -5.86 & 1.00 \\
    	\cline{3-5}
    	& & \multirow{1}{*}{[SII]~4068}
    		& -4.43 & 1.12 \\
    	\cline{3-5}
    	& & \multirow{1}{*}{[SII]~6730}
    		& -4.65 & 1.17 \\
    \hline
    \multirow{4}{*}{0.31}
    & \multirow{4}{*}{0.031}
    	& \multirow{1}{*}{[OI]~6300}
    		& -4.22 & 1.26 \\
    	\cline{3-5}
    	& & \multirow{1}{*}{[OI]~5577}
    		& -5.35 & 1.02 \\
    	\cline{3-5}
    	& & \multirow{1}{*}{[SII]~4068}
    		& -3.70 & 1.41 \\
    	\cline{3-5}
    	& & \multirow{1}{*}{[SII]~6730}
    		& -4.16 & 1.48 \\
    \hline
    \multirow{4}{*}{0.31}
    & \multirow{4}{*}{0.1}
    	& \multirow{1}{*}{[OI]~6300}
    		& -4.09 & 1.70 \\
    	\cline{3-5}
    	& & \multirow{1}{*}{[OI]~5577}
    		& -5.31 & 1.12 \\
    	\cline{3-5}
    	& & \multirow{1}{*}{[SII]~4068}
    		& -3.49 & 2.29 \\
    	\cline{3-5}
    	& & \multirow{1}{*}{[SII]~6730}
    		& -3.96 & 2.34 \\
\end{tabular}
    \caption{Total line luminosities from the photoevaporative wind models with an additional EUV jet source at a height of 50 AU on the z-axis that is modelled as a blackbody with T = 25000 K with a luminosity that is given in the table as L$_{\mathrm{jet}}$. The last column shows the ratios of the luminosities between a model with and without the jet source.}
    \label{tab:luminosities_pe_jet}
\end{table}

\begin{table}
    \centering
    \begin{tabular}{ccrrrcr}
    \multicolumn{1}{l}{L$_{\mathrm{acc}}$} & \multicolumn{1}{l}{L$_{\mathrm{jet}}$}
        & \multicolumn{1}{c}{i} & \multicolumn{1}{c}{v$_\mathrm{c}$} & \multicolumn{1}{c}{FWHM} & \multirow{2}{*}{log\big($\mathrm{\tfrac{L_{comp}}{L_{\odot}}}$\big)} & \multicolumn{1}{c}{\multirow{2}{*}{Type}} \\\relax
         [L$_{\odot}$] & [L$_{\odot}$] & [\degree] 
        & \multicolumn{1}{c}{[km/s]} & \multicolumn{1}{c}{[km/s]} & & \\
    \hline
    \hline
    \multirow{5}{*}{\rotatebox[origin=c]{90}{$2.6\cdot10^{-2}$}}
    & \multirow{5}{*}{\rotatebox[origin=c]{90}{$2.6\cdot10^{-3}$}}
    	& \multirow{1}{*}{0}
    		& -3.16 & 19.64 & -4.76 & NLVC\\
    	\cline{3-7}
    	& & \multirow{1}{*}{20}
    		& -4.62 & 25.33 & -4.75 & NLVC\\
    	\cline{3-7}
    	& & \multirow{1}{*}{40}
    		& -5.53 & 31.43 & -4.75 & NLVC\\
    	\cline{3-7}
    	& & \multirow{1}{*}{60}
    		& -5.84 & 29.63 & -4.94 & NLVC\\
    	\cline{3-7}
    	& & \multirow{1}{*}{80}
    		& -2.52 & 24.32 & -5.11 & NLVC\\
    \hline
    \multirow{5}{*}{\rotatebox[origin=c]{90}{0.31}}
    & \multirow{5}{*}{\rotatebox[origin=c]{90}{0.031}}
    	& \multirow{1}{*}{0}
    		& -6.59 & 23.10 & -4.21 & NLVC\\
    	\cline{3-7}
    	& & \multirow{1}{*}{20}
    		& -7.35 & 24.42 & -4.21 & NLVC\\
    	\cline{3-7}
    	& & \multirow{1}{*}{40}
    		& -7.55 & 25.29 & -4.23 & NLVC\\
    	\cline{3-7}
    	& & \multirow{1}{*}{60}
    		& -6.62 & 23.21 & -4.35 & NLVC\\
    	\cline{3-7}
    	& & \multirow{1}{*}{80}
    		& -2.60 & 20.90 & -4.45 & NLVC\\
    \hline
    \multirow{5}{*}{\rotatebox[origin=c]{90}{0.31}}
    & \multirow{5}{*}{\rotatebox[origin=c]{90}{0.1}}
    	& \multirow{1}{*}{0}
    		& -9.88 & 26.04 & -4.09 & NLVC\\
    	\cline{3-7}
    	& & \multirow{1}{*}{20}
    		& -10.26 & 23.74 & -4.09 & NLVC\\
    	\cline{3-7}
    	& & \multirow{1}{*}{40}
    		& -9.75 & 22.17 & -4.11 & NLVC\\
    	\cline{3-7}
    	& & \multirow{1}{*}{60}
    		& -7.49 & 20.39 & -4.20 & NLVC\\
    	\cline{3-7}
    	& & \multirow{1}{*}{80}
    		& -2.74 & 19.36 & -4.26 & NLVC\\

\end{tabular}
    \caption{Centroid velocities, FWHM, luminosity and type of the profile components from the photoevaporative wind models with an additional EUV jet source at a height of 50 AU on the z-axis that is modelled as a blackbody with T = 25000 K with a luminosity that is given in the table as L$_{\mathrm{jet}}$.}
    \label{tab:fits_pe_jet}
\end{table}

\section{Discussion}\label{sec:discussion}
\subsection{Origin of the correlations with L\texorpdfstring{$_{\mathrm{acc}}$}{}} \label{sec:discussion:correlations}

We have shown in section \ref{sec:res:correlations_acc} that the photoevaporation model has the same correlations with the accretion luminosities as observed NLVCs. However BLVCs are not reproduced in a photoevaporative wind, therefore we have shown that two different wind types can produce the same correlations. To explain why these correlations exist we can again use the emission maps in Figure \ref{fig:nh_v_emission_all}: In the photoevaporation model a higher accretion luminosity increases the heated wind region and thus the emission region. This has not only the consequence of higher luminosities, but also more emission from the upper, faster regions of the wind, increasing the blueshift of the line. At the same time the line width is reduced, because the azimuthal velocity is very low in the upper wind regions and the Kepler broadening is reduced. The top right panel of figure \ref{fig:corr_acc} shows that this effect is strongest for the highest inclinations. The 0\degree{} profile shows exactly the opposite behaviour, because Kepler rotation is not at play and the broadness is entirely dominated by the velocity gradient, which is greater in a larger emission region. 

In the case of an MHD wind the increase in size of the emission region with higher L$_{\mathrm{acc}}$ is less pronounced, because it is countered by the increase in gas density. As a result, there is a strong correlation between accretion luminosities and line luminosity, but the line profiles remain similar. An exception is the model with the lowest accretion luminosity where the inner edge of the wind is highly ionized such that no flux with a blueshift > 100~km/s can be observed, as discussed in detail in section \ref{sec:res:emission_regions}. With similar profiles, the correlation between the profile components and L$_{\mathrm{acc}}$ are not as obvious as in the observations. This could be an indication that our model overestimates the link between wind mass-loss and accretion rate. With a weaker correlation we would expect a bigger increase in the size of the emission regions and consequently correlations that are similar to those found in the photoevaporation models and the observations. Alternatively, as proposed by \citetalias{Banzatti2018}, an MHD wind where the increase in mass-loss manifests not only in a higher gas density but also in higher outflow velocities could restore the correlations.

\subsection{Correlations with viewing angle}
\label{sec:discussion:corr_inc}

\citetalias{Banzatti2018} interpret the observed correlation between NLVC line widths and viewing angle and the lack of such a correlation in the BLVC as possible evidence that the NLVCs are emitted close to the disc surface at larger radii, rotating close to the Keplerian speed, while the BLVCs are emitted at higher parts of the wind, where the poloidal velocity is increased and the toroidal velocity is lower than the Keplerian speed at the footpoint where the flow was launched. As discussed in section \ref{sec:res:profile_decomposition} the Gaussian decomposition and with it the correlations of their properties should only be treated with caution, but we can nevertheless investigate the plausibility of this scenario: In our MHD wind models with higher accretion luminosities the emission region of the flux with zero blueshift is indeed located at a height of $\gtrsim$ 2 au, but this includes the narrow component as well. The scenario would thus only be plausible if the MHD profile was supplemented by another narrow low velocity component produced by a different wind at larger radii and closer to the disc surface. Evidence for such a combination of profiles is the overabundance of Keplerian double peaks in our models. Double peaks are rarely observed, which means that either our model overestimates azimuthal wind velocities or in reality the Keplerian throughts are filled in by another component.

\subsection{Combination of MHD and photoevaporation profiles} \label{sec:discussion:profiles_combi}

 One possible solution for both problems described in the previous section is the combination of a higher-density inner MHD wind and a photoevaporative wind that dominates at larger radii and fills the Kepler throught of the MHD profile. If the MHD wind arises from a region that is MRI turbulent we would expect the wind itself to become turbulent and introduce asymmetries that allow the X-ray to reach the disc at high-enough radii. Furthermore, knots/ Herbig-Haro objects in jets are evidence for a temporal variability of the outflow velocity. It might be plausible to assume a (possibly time dependent) filling factor to modulate the amount of radiation finally reaching the disc at radii where it can drive a photoevaporative wind.
 
 \begin{figure}
    \centering
    \includegraphics[width=.48\textwidth]{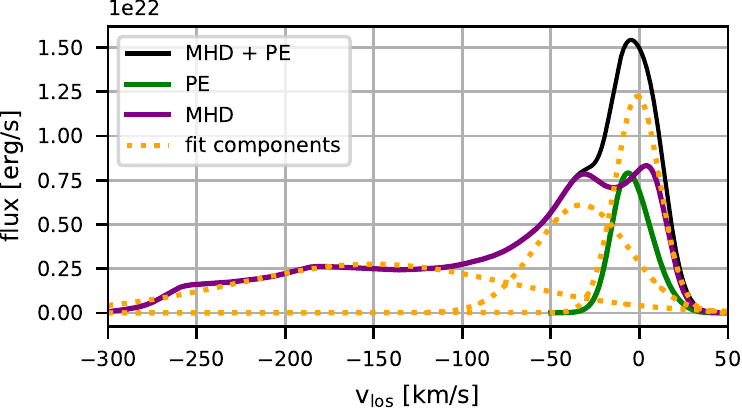}
    \caption{Combination of MHD and photoevaporative wind profile for the [OI]~6300 line of our models with with L$_{\mathrm{acc}}$ = 0.31~L$_{\odot}$ at 40\degree{} inclination.}
    \label{fig:profile_combi}
\end{figure}
 
\begin{figure*}
    \centering
    \includegraphics[width=.9\textwidth]{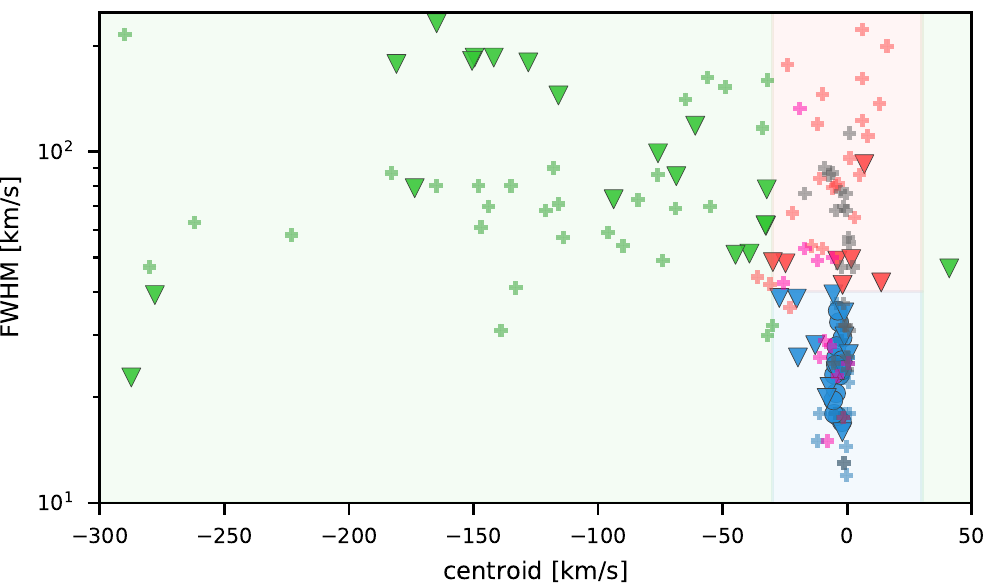}
    \caption{Overview of the parameter space of centroid velocities and FWHM of the components of the combined [OI]~6300 MHD + photoevaporation profiles. Components from the MHD+photoevaporation combination models are marked with a triangle, those from the photoevaporation model with a circle. The plusses mark observations by \citetalias{Banzatti2018}. The colors indicate the type of the components: Green: HVC, red: BLVC, blue: NLVC, grey: SC, pink: SCJ}
    \label{fig:OI-6300_overview_combi}
\end{figure*}

We performed this experiment by combining our MHD with the photoevaporation profiles to see whether this could indeed bring the total profiles closer to the observations. For this purpose, we added a fraction of the flux of the photoevaporative profiles to that of the MHD profile. We chose fractions of 24~\%, 12~\% and 6~\% for the models with the lowest, intermediate and highest accretion luminosity, respectively. It is important to note that this choice is chosen only because it provided acceptable results and has no further justification, besides our expectation that a denser inner MHD wind, which is the case in the models with higher L$_{\mathrm{acc}}$, will shield more of the X-ray radiation that drives the photoevaporative wind. Figure \ref{fig:profile_combi} shows the combined [OI]~6300 profile at 40\degree{} inclination for the model with L$_{\mathrm{acc}}$ = 0.31~L$_{\odot}$. The Keplerian double peaks are indeed filled by the photoevaporation profile. Note that the fits do not properly separate the photoevaporative from the MHD component. Figure \ref{fig:OI-6300_overview_combi} shows that combining the profiles does improve the centroid and FWHM distribution of the modelled profiles. Especially the number of highly blue- or redshifted NLVCs is greatly reduced and the number of BLVCs increased from 4 to 7. The BLVCs with high FHWM $\gtrsim$ 100~km/s are still not reproduced, however, it is reassuring that our simple experiment can already significantly improve the distribution of component properties.
 
\subsection{Model limitations} \label{sec:discussion:limitations}

Our MHD model, while to a good degree consistent with observations, uses many simplifying assumptions. Most obviously, it lacks a thick disc which will certainly affect the velocity and density structure of the wind at larger radii. This could significantly affect the line profiles, especially in the case of high accretion luminosities, when the emission regions reach farther into that area. Moreover we have seen that the model tends to underestimate the line luminosities and to overestimate the outflow velocities despite the relatively large inner radius. Detailed numerical models with much lower inner radii and lower floor densities than existing models would be valuable to accurately calculate synthetic line profiles from MHD winds. Models that combine MHD with photoevaporation with an adjustable ratio between them, such as the model by \citet{Rodenkirch2019a} could be particularly useful to determine the relative importance of MHD and photoevaporative winds.

Our synthetic line profiles have all been calculated from wind models with fixed parameters, with the luminosity of the illuminating UV component being the only variable. Although \citet{Ercolano2010} have shown that the X-ray luminosity has no effect on the luminosities of the collisionally excited lines, it does have an effect on the photoevaporative wind mass-loss rates in the sense that a star with a higher X-ray luminosity is able to drive more rigorous winds \citep{Picogna2019}. Similarly, the stellar mass can affect the mass-loss rates of a photoevaporative wind (Picogna et al., in preparation). In a more extensive study that includes multiple wind models calculated with different initial conditions, we would expect the synthetic profile components to cover an even larger parameter space of observations.


\section{Conclusions}\label{sec:conclusions}
We have calculated the photoionization structure of a detailed numerical X-ray photoevaporative disc wind model and a simple magnetocentrifugally driven wind model and created 2-dimensional emission maps of the [OI] 6300, [OI] 5577, [SII] 4068 and [SII] 6730 lines. We investigated the location and physical properties of their emission regions and studied the influence of the accretion luminosity as the dominating heating source of the wind. We used the 2D emission maps to construct 3-dimensional models of the line emission in axisymmetric disc winds and calculated theoretical spectra of the lines as they would be expected from observations at different viewing angles. In these line profile calculations we accounted for Keplerian broadening, thermal broadening, dust absorption and a limited spectral resolution. We followed the commonly used approach of decomposing our synthetic line profiles into multiple Gaussian components to assess the informative value that is held by the individual components and to explore the correlations between the component properties and the disc inclination and accretion luminosity in comparison to observations. Our main findings are:

\begin{itemize}
	\item The X-ray photoevaporative disc wind model can successfully reproduce simple, single-Gaussian lines with small blue shifts (2--5~km/s) and FWHM between 15 and 40~km/s that overlap well with the observations of NLVCs. The model cannot reproduce, however, the observed NLVCs with the smallest and largest blueshift near 0 and 12~km/s and the narrowest line width near 10~km/s. Neither can it reproduce any of the observed BLVCs or HVCs and therefore needs to be coupled to another wind type in order to fully explain the line profiles observed in full/primordial discs.
	\item The MHD wind model on its own can reproduce complex line profiles that can be decomposed into all component types (NLVC, BLVC and even HVC). The kinematic properties of the components approximately resemble those of observations but the profiles often show Keplerian double peaks that are rarely observed. These double peaks affect the profile decomposition (NLVCs with blue- and redshifts near 20~km/s are often produced), making it difficult to draw convincing conclusions on the correlations of component properties. The correlations of the component luminosities with the accretion luminosity are well reproduced, but not the correlations of their centroid velocities or FWHMs, except for the HVCs that vaguely follow the observed trend. The model cannot produce the narrowest NLVCs with FWHM < 20~km/s and BLVCs are produced rarely and only with line widths < 100~km/s.
	\item A combination of MHD and photoevaporation profiles is plausible and could alleviate the problems that are caused by the Keplerian double peaks. Wind asymmetries and time variability might allow the radiation to reach and heat the disc at radii where a photoevaporative wind can be launched. To test this scenario a model combining MHD and photoevaporative winds is required, but well beyond the scope of this paper. Currently available MHD calculations lack the required resolution to model forbidden line emission in the wind.  
	\item An additional EUV heating source located at higher heights in a jet along the z-axis can significantly increase the luminosity and blueshifts of the profiles in a photoevaporative wind. The [OI]~6300 line luminosity was increased by up to 70~\% and its blueshift could reach up to 10~km/s, close to the fastest observed NLVC blueshifts near 12~km/s.
	\item In reference to recent observational studies, this work shows that the fitted Gaussian components do not clearly separate the flux from physically distinct emission regions and that the line broadening is often determined more by the velocity gradient in the wind than by Keplerian rotation in the disc, and as such line widths are not good tracers of the radial disk region where the wind is launched. This confirms recent analyses by \citet{Banzatti2018}, where cautionary notes on using line widths to infer emitting radii were proposed, based on the absence of correlations with the viewing angle (Sect. 5.1 in \citetalias{Banzatti2018}). Future observational works should be cautious as well, and consider the impact of the velocity gradient in the wind.
	\item The EUV flux that is dominated by the accretion luminosity has a strong effect on the size of the emission region. A higher luminosity can heat a larger region of the wind to the temperatures required for the collisionally excited lines to be emitted. This leads to correlations between profile component properties and the accretion luminosity that are positive for the component luminosities and centroid velocities and is negative for the FWHM. This effect is weaker for our MHD wind models, because the expected increase in mass-loss with increasing accretion luminosity counteracts the growth of the emission region. This could be different in other MHD models where also the outflow velocities increase with increasing mass-loss rate. 
	\item The observed connections between NLVC and BLVC properties could be explained by the correlations with the accretion luminosity. Our X-ray photoevaporation model can successfully reproduce these correlations for the NLVCs and is thus consistent with a scenario where the NLVCs are launched by a photoevaporative wind, while the BLVCs trace a wind that is driven by a different mechanism. 
\end{itemize}

\section*{Acknowledgements}\label{sec:acknowledgements}
We thank the anonymous referee for helpful comments on the first version of this paper. We thank Ilaria Pascucci for providing the [OI] 6300 line profile of DG Tau and for insightful discussions. We thank Manuel G\"udel for the helpful discussion about DG Tau. GP acknowledges support from the DFG Research Unit "Transition Disks" (FOR 2634/1, ER 685/8-1). BE acknowledges support from the DFG Research Unit "Transition Disks" (FOR 2634/1, ER 685/11-1).




\bibliographystyle{mnras}
\bibliography{mendeley}



\appendix
\section{Correlation analysis of NLVC and BLVC component properties with L\texorpdfstring{$_{\mathrm{acc}}$}{}} \label{sec:appendix:correlations}

\begin{figure*}
    \centering
    \includegraphics[width=.9\textwidth]{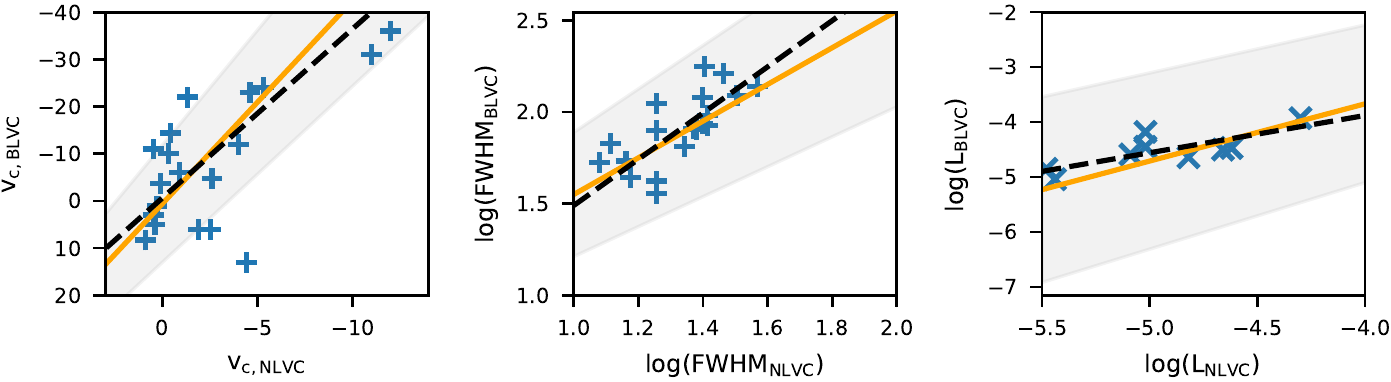}
    \caption{Black dashed lines: correlations between the component properties. Orange solid lines: correlations as we would expect them if they were solely a result of the correlations with the accretion luminosity. The correlations and observations in the left and middle panels are those reported by \citetalias{Banzatti2018}, the observations in the right panel by \citet{Fang2018}. The black dashed line in the right panel was obtained in this work as described in appendix \ref{sec:appendix:correlations}.}
    \label{fig:expected_corr_comps}
\end{figure*}

As described in section \ref{sec:results}, \citetalias{Banzatti2018} have found that the centroid velocities, FWHM and equivalent widths of the NLVCs correlate with those of the BLVCs. They also found that the centroid velocities and FWHM of both components have similar correlations with the accretion luminosity and another correlation between component luminosity and accretion luminosity is well known (section \ref{sec:introduction}). We will show here that the correlations between NLVC properties and L$_{\mathrm{acc}}$ and between BLVC properties and L$_{\mathrm{acc}}$ are indeed sufficient to explain the correlation between BLVC and NLVC properties. If the component property $y$ is correlated with the accretion luminosity L$_{\mathrm{acc}}$ via
\begin{equation}
    y_{\mathrm{BLVC}} = a + b ~log\big(\mathrm{L_{acc}}\big)
\end{equation}
and
\begin{equation}
    y_{\mathrm{NLVC}} = c + d ~log\big(\mathrm{L_{acc}}\big)
\end{equation}
and if we assume that $y$ is not correlated with another variable, we would expect that
\begin{equation}
    y_{\mathrm{BLVC}} = a - \frac{b}{d} \big(c + y_{\mathrm{NLVC}}\big).
\end{equation}

\begin{table}
    \centering
    \begin{tabular}{llrrc}
    \multicolumn{1}{c}{y} & \multicolumn{1}{c}{x} & \multicolumn{1}{c}{a} & \multicolumn{1}{c}{b} & Ref. \\
    \hline
    \hline
    v$_{\mathrm{c,NLVC}}$ & log L$_{\mathrm{acc}}$ & -6.20 $\pm$ 1.10 & -3.70 $\pm$ 0.80 & 1 \\
    v$_{\mathrm{c,BLVC}}$ & log L$_{\mathrm{acc}}$ & -26.10 $\pm$ 4.50 & -15.90 $\pm$ 3.20 & 1 \\
    \hline
    log FWHM$_{\mathrm{NLVC}}$ & log L$_{\mathrm{acc}}$ & 0.98 $\pm$ 0.07 & -0.34 $\pm$ 0.05 & 1 \\
    log FWHM$_{\mathrm{BLVC}}$ & log L$_{\mathrm{acc}}$ & 1.53 $\pm$ 0.08 & -0.34 $\pm$ 0.06 & 1 \\
    \hline
    log L$_{\mathrm{NLVC}}$ & log L$_{\mathrm{acc}}$ & -4.53 $\pm$ 0.14 & 0.49 $\pm$ 0.10 & 2 \\
    log L$_{\mathrm{BLVC}}$ & log L$_{\mathrm{acc}}$ & -4.22 $\pm$ 0.07 & 0.51 $\pm$ 0.05 & 2 \\
    \hline
    \multicolumn{2}{l}{1: \citet{Banzatti2018}} & \\
    \multicolumn{2}{l}{2: \citet{Fang2018}} & \\
\end{tabular}
    \caption{Fit parameters for the correlations of the centroids, FWHMs and component luminosities with the accretion luminosity and for the correlations between the LVC properties.}
    \label{tab:fits_corr_lacc}
\end{table}

We tested this expectation against the observed correlations using the fit parameters listed in table \ref{tab:fits_corr_lacc}. Since \citetalias{Banzatti2018} do not report fit parameters for the correlation between component luminosity and accretion luminosity we use the parameters reported by \citet{Fang2018} and perform a fit to the BLVC - NLVC luminosity correlation in their sample to test the connection between the component luminosities. The relation we found is 
\begin{equation}
    log\big(\mathrm{L_{BLVC}}\big) = -6.48 + 4.31 log\big(\mathrm{L_{NLVC}}\big).
\end{equation}
Errors are not stated, because we lack the uncertainties of the underlying data, but for the purpose of this test we can safely ignore them. The results are shown in figure \ref{fig:expected_corr_comps}, which shows that the connections between NLVC and BLVC properties can indeed be explained by their correlations with the accretion luminosity.

\section{Complementary figures and tables}\label{sec:appendix:fits}

\begin{table*}
    \centering
    \begin{tabular}{crrrcrrrcrrrcr}
    & & \multicolumn{4}{c}{L$_{\mathrm{acc}}$ = $2.6\cdot10^{-2}$ L$_{\odot}$}
         & \multicolumn{4}{c}{L$_{\mathrm{acc}}$ = 0.31 L$_{\odot}$}
         & \multicolumn{4}{c}{L$_{\mathrm{acc}}$ = 1 L$_{\odot}$} \\ [1.5ex]
    \multirow{2}{*}{Line} & \multicolumn{1}{c}{i} 
        & \multicolumn{1}{c}{v$_\mathrm{c}$} & \multicolumn{1}{c}{FWHM} & \multirow{2}{*}{log\big($\mathrm{\tfrac{L_{comp}}{L_{\odot}}}$\big)} & \multicolumn{1}{c}{\multirow{2}{*}{Type}}
        & \multicolumn{1}{c}{v$_\mathrm{c}$} & \multicolumn{1}{c}{FWHM} & \multirow{2}{*}{log\big($\mathrm{\tfrac{L_{comp}}{L_{\odot}}}$\big)} & \multicolumn{1}{c}{\multirow{2}{*}{Type}}
        & \multicolumn{1}{c}{v$_\mathrm{c}$} & \multicolumn{1}{c}{FWHM} & \multirow{2}{*}{log\big($\mathrm{\tfrac{L_{comp}}{L_{\odot}}}$\big)} & \multicolumn{1}{c}{\multirow{2}{*}{Type}} \\
    & [\degree] 
        & \multicolumn{1}{c}{[km/s]} & \multicolumn{1}{c}{[km/s]} & & 
        & \multicolumn{1}{c}{[km/s]} & \multicolumn{1}{c}{[km/s]} & & 
        & \multicolumn{1}{c}{[km/s]} & \multicolumn{1}{c}{[km/s]} & &  \\
    \hline
    \hline
    \multirow{5}{*}{\rotatebox[origin=c]{90}{[OI] 6300}}
    	& \multirow{1}{*}{0}
    		& -2.11 & 16.98 & -4.82 & NLVC	 & 	-3.95 & 17.68 & -4.32 & NLVC	 & 	-5.27 & 17.93 & -3.90 & NLVC\\
    	\cline{2-14}
    	& \multirow{1}{*}{20}
    		& -2.95 & 23.08 & -4.82 & NLVC	 & 	-4.49 & 20.54 & -4.32 & NLVC	 & 	-5.58 & 19.60 & -3.90 & NLVC\\
    	\cline{2-14}
    	& \multirow{1}{*}{40}
    		& -3.28 & 32.72 & -4.82 & NLVC	 & 	-4.48 & 25.95 & -4.33 & NLVC	 & 	-5.25 & 23.20 & -3.91 & NLVC\\
    	\cline{2-14}
    	& \multirow{1}{*}{60}
    		& -3.79 & 35.28 & -5.03 & NLVC	 & 	-4.30 & 27.95 & -4.49 & NLVC	 & 	-4.56 & 24.76 & -4.04 & NLVC\\
    	\cline{2-14}
    	& \multirow{1}{*}{80}
    		& -1.88 & 29.53 & -5.28 & NLVC	 & 	-2.11 & 25.49 & -4.66 & NLVC	 & 	-2.08 & 23.87 & -4.13 & NLVC\\
    \hline
    \multirow{9}{*}{\rotatebox[origin=c]{90}{[OI] 5577}}
    	& \multirow{1}{*}{0}
    		& -0.45 & 14.98 & -5.87 & NLVC	 & 	-0.64 & 15.37 & -5.36 & NLVC	 & 	-0.92 & 15.77 & -4.94 & NLVC\\
    	\cline{2-14}
    	& \multirow{1}{*}{20}
    		& -0.97 & 31.39 & -5.88 & NLVC	 & 	-1.09 & 27.31 & -5.36 & NLVC	 & 	-1.56 & 25.40 & -4.94 & NLVC\\
    	\cline{2-14}
    	& \multirow{2}{*}{40}
    		& 14.79 & 26.97 & -6.19 & NLVC	 & 	13.95 & 24.55 & -5.75 & NLVC	 & 	-1.93 & 40.90 & -4.95 & BLVC\\
    	&	& -16.92 & 26.12 & -6.18 & NLVC	 & 	-11.96 & 27.69 & -5.60 & NLVC	 & 	 &  &  & \\
    	\cline{2-14}
    	& \multirow{3}{*}{60}
    		& 26.60 & 24.14 & -6.86 & NLVC	 & 	20.10 & 29.79 & -6.26 & NLVC	 & 	-3.76 & 46.22 & -5.31 & BLVC\\
    	&	& -6.92 & 39.08 & -6.61 & NLVC	 & 	-12.49 & 37.38 & -5.91 & NLVC	 & 	 &  &  & \\
    	&	& -30.11 & 19.92 & -6.97 & HVC	 & 	 &  &  & 	 & 	 &  &  & \\
    	\cline{2-14}
    	& \multirow{2}{*}{80}
    		& 8.98 & 20.32 & -7.95 & NLVC	 & 	-2.52 & 29.43 & -6.56 & NLVC	 & 	-2.55 & 26.87 & -5.94 & NLVC\\
    	&	& -11.44 & 19.09 & -7.90 & NLVC	 & 	 &  &  & 	 & 	 &  &  & \\
    \hline
    \multirow{5}{*}{\rotatebox[origin=c]{90}{[SII] 4068}}
    	& \multirow{1}{*}{0}
    		& -2.34 & 16.27 & -4.48 & NLVC	 & 	-3.83 & 16.81 & -3.85 & NLVC	 & 	-4.89 & 17.45 & -3.47 & NLVC\\
    	\cline{2-14}
    	& \multirow{1}{*}{20}
    		& -3.09 & 21.32 & -4.48 & NLVC	 & 	-4.24 & 19.17 & -3.85 & NLVC	 & 	-5.13 & 18.84 & -3.47 & NLVC\\
    	\cline{2-14}
    	& \multirow{1}{*}{40}
    		& -3.38 & 29.93 & -4.49 & NLVC	 & 	-4.11 & 23.85 & -3.86 & NLVC	 & 	-4.77 & 21.88 & -3.48 & NLVC\\
    	\cline{2-14}
    	& \multirow{1}{*}{60}
    		& -4.88 & 27.03 & -4.98 & NLVC	 & 	-4.55 & 23.66 & -4.18 & NLVC	 & 	-4.59 & 21.72 & -3.71 & NLVC\\
    	\cline{2-14}
    	& \multirow{1}{*}{80}
    		& -2.19 & 26.57 & -5.04 & NLVC	 & 	-1.91 & 24.74 & -4.16 & NLVC	 & 	-1.76 & 22.75 & -3.68 & NLVC\\
    \hline
    \multirow{5}{*}{\rotatebox[origin=c]{90}{[SII] 6730}}
    	& \multirow{1}{*}{0}
    		& -7.54 & 18.26 & -4.70 & NLVC	 & 	-7.81 & 18.84 & -4.33 & NLVC	 & 	-8.13 & 19.59 & -4.05 & NLVC\\
    	\cline{2-14}
    	& \multirow{1}{*}{20}
    		& -7.28 & 18.47 & -4.70 & NLVC	 & 	-7.47 & 18.88 & -4.32 & NLVC	 & 	-7.75 & 19.45 & -4.05 & NLVC\\
    	\cline{2-14}
    	& \multirow{1}{*}{40}
    		& -6.20 & 19.17 & -4.70 & NLVC	 & 	-6.29 & 19.09 & -4.33 & NLVC	 & 	-6.49 & 19.23 & -4.05 & NLVC\\
    	\cline{2-14}
    	& \multirow{1}{*}{60}
    		& -4.37 & 19.54 & -4.74 & NLVC	 & 	-4.38 & 19.03 & -4.35 & NLVC	 & 	-4.48 & 18.77 & -4.07 & NLVC\\
    	\cline{2-14}
    	& \multirow{1}{*}{80}
    		& -1.51 & 19.62 & -4.74 & NLVC	 & 	-1.49 & 18.97 & -4.35 & NLVC	 & 	-1.51 & 18.49 & -4.07 & NLVC\\
\end{tabular}
    \caption{Centroid velocities, FWHM, luminosity and type of the individual components obtained by fitting the profiles from the photoevaporative wind models.}
    \label{tab:fits_pe}
\end{table*}

\begin{table*}
    \centering
    \begin{tabular}{crrrcrrrcrrrcr}
    & & \multicolumn{4}{c}{L$_{\mathrm{acc}}$ = $2.6\cdot10^{-2}$ L$_{\odot}$}
         & \multicolumn{4}{c}{L$_{\mathrm{acc}}$ = 0.31 L$_{\odot}$}
         & \multicolumn{4}{c}{L$_{\mathrm{acc}}$ = 1 L$_{\odot}$} \\ [1.5ex]
    \multirow{2}{*}{Line} & \multicolumn{1}{c}{i} 
        & \multicolumn{1}{c}{v$_\mathrm{c}$} & \multicolumn{1}{c}{FWHM} & \multirow{2}{*}{log\big($\mathrm{\tfrac{L_{comp}}{L_{\odot}}}$\big)} & \multicolumn{1}{c}{\multirow{2}{*}{Type}}
        & \multicolumn{1}{c}{v$_\mathrm{c}$} & \multicolumn{1}{c}{FWHM} & \multirow{2}{*}{log\big($\mathrm{\tfrac{L_{comp}}{L_{\odot}}}$\big)} & \multicolumn{1}{c}{\multirow{2}{*}{Type}}
        & \multicolumn{1}{c}{v$_\mathrm{c}$} & \multicolumn{1}{c}{FWHM} & \multirow{2}{*}{log\big($\mathrm{\tfrac{L_{comp}}{L_{\odot}}}$\big)} & \multicolumn{1}{c}{\multirow{2}{*}{Type}} \\
    & [\degree] 
        & \multicolumn{1}{c}{[km/s]} & \multicolumn{1}{c}{[km/s]} & & 
        & \multicolumn{1}{c}{[km/s]} & \multicolumn{1}{c}{[km/s]} & & 
        & \multicolumn{1}{c}{[km/s]} & \multicolumn{1}{c}{[km/s]} & &  \\
    \hline
    \hline
    \multirow{17}{*}{\rotatebox[origin=c]{90}{[OI] 6300}}
    	& \multirow{4}{*}{0}
    		& -19.03 & 19.92 & -5.91 & NLVC	 & 	-19.46 & 20.17 & -5.16 & NLVC	 & 	-21.43 & 21.69 & -4.52 & NLVC\\
    	&	& -33.97 & 33.16 & -5.81 & HVC	 & 	-41.10 & 43.33 & -5.17 & HVC	 & 	-44.89 & 44.99 & -4.51 & HVC\\
    	&	&  &  &  & 	 & 	-144.63 & 184.61 & -4.90 & HVC	 & 	-140.47 & 180.04 & -4.17 & HVC\\
    	&	&  &  &  & 	 & 	-277.31 & 40.99 & -5.49 & HVC	 & 	-286.89 & 24.39 & -5.14 & HVC\\
    	\cline{2-14}
    	& \multirow{4}{*}{20}
    		& -17.09 & 28.40 & -5.80 & NLVC	 & 	-17.57 & 32.49 & -5.08 & NLVC	 & 	-21.96 & 36.94 & -4.37 & NLVC\\
    	&	& -36.95 & 42.90 & -5.91 & HVC	 & 	-45.06 & 53.66 & -5.21 & HVC	 & 	-59.66 & 71.97 & -4.49 & HVC\\
    	&	&  &  &  & 	 & 	-129.69 & 138.55 & -5.03 & HVC	 & 	-168.16 & 187.85 & -4.22 & HVC\\
    	&	&  &  &  & 	 & 	-251.53 & 75.66 & -5.30 & HVC	 & 	 &  &  & \\
    	\cline{2-14}
    	& \multirow{3}{*}{40}
    		& 2.22 & 20.68 & -6.27 & NLVC	 & 	5.67 & 21.10 & -5.53 & NLVC	 & 	4.12 & 21.95 & -4.88 & NLVC\\
    	&	& -25.83 & 55.91 & -5.65 & BLVC	 & 	-28.51 & 61.77 & -4.94 & BLVC	 & 	-31.40 & 62.53 & -4.29 & HVC\\
    	&	&  &  &  & 	 & 	-147.60 & 185.32 & -4.84 & HVC	 & 	-127.46 & 180.09 & -4.16 & HVC\\
    	\cline{2-14}
    	& \multirow{3}{*}{60}
    		& 14.59 & 27.48 & -6.23 & NLVC	 & 	21.35 & 27.84 & -5.67 & NLVC	 & 	20.47 & 29.27 & -4.91 & NLVC\\
    	&	& -30.12 & 56.32 & -5.93 & HVC	 & 	-34.92 & 98.13 & -5.00 & HVC	 & 	-31.19 & 82.48 & -4.43 & HVC\\
    	&	&  &  &  & 	 & 	-145.40 & 117.85 & -5.22 & HVC	 & 	-118.13 & 143.92 & -4.44 & HVC\\
    	\cline{2-14}
    	& \multirow{3}{*}{80}
    		& 20.97 & 32.05 & -6.38 & NLVC	 & 	23.53 & 71.80 & -5.44 & BLVC	 & 	36.95 & 54.87 & -4.84 & HVC\\
    	&	& -26.41 & 44.77 & -6.24 & BLVC	 & 	-63.04 & 103.46 & -5.33 & HVC	 & 	-38.01 & 103.82 & -4.63 & HVC\\
    	&	&  &  &  & 	 & 	-107.95 & 41.69 & -5.96 & HVC	 & 	-102.76 & 59.58 & -4.97 & HVC\\
    \hline
    \multirow{15}{*}{\rotatebox[origin=c]{90}{[OI] 5577}}
    	& \multirow{3}{*}{0}
    		& -32.46 & 34.03 & -7.10 & HVC	 & 	-23.80 & 23.24 & -6.04 & NLVC	 & 	-23.98 & 23.52 & -5.20 & NLVC\\
    	&	&  &  &  & 	 & 	-46.96 & 45.96 & -6.10 & HVC	 & 	-48.35 & 48.28 & -5.23 & HVC\\
    	&	&  &  &  & 	 & 	-110.03 & 136.64 & -6.18 & HVC	 & 	-118.27 & 148.64 & -5.27 & HVC\\
    	\cline{2-14}
    	& \multirow{3}{*}{20}
    		& -12.04 & 26.94 & -7.74 & NLVC	 & 	-13.71 & 30.35 & -6.26 & NLVC	 & 	-17.41 & 34.36 & -5.28 & NLVC\\
    	&	& -40.64 & 53.19 & -7.21 & HVC	 & 	-40.01 & 52.69 & -5.98 & HVC	 & 	-45.07 & 56.09 & -5.17 & HVC\\
    	&	&  &  &  & 	 & 	-106.49 & 142.98 & -6.16 & HVC	 & 	-118.83 & 153.12 & -5.28 & HVC\\
    	\cline{2-14}
    	& \multirow{3}{*}{40}
    		& 9.38 & 34.09 & -7.64 & NLVC	 & 	10.85 & 26.68 & -6.38 & NLVC	 & 	9.07 & 25.87 & -5.53 & NLVC\\
    	&	& -46.55 & 77.04 & -7.25 & HVC	 & 	-34.86 & 78.33 & -5.89 & HVC	 & 	-34.06 & 74.00 & -5.04 & HVC\\
    	&	&  &  &  & 	 & 	-116.18 & 162.23 & -6.23 & HVC	 & 	-111.72 & 157.53 & -5.30 & HVC\\
    	\cline{2-14}
    	& \multirow{3}{*}{60}
    		& 28.18 & 52.77 & -7.88 & BLVC	 & 	32.92 & 48.76 & -6.55 & HVC	 & 	29.17 & 44.80 & -5.67 & BLVC\\
    	&	& -53.82 & 81.58 & -7.70 & HVC	 & 	-42.17 & 89.84 & -6.34 & HVC	 & 	-40.38 & 81.32 & -5.51 & HVC\\
    	&	&  &  &  & 	 & 	-131.76 & 165.80 & -6.72 & HVC	 & 	-110.93 & 174.87 & -5.70 & HVC\\
    	\cline{2-14}
    	& \multirow{3}{*}{80}
    		& 26.95 & 45.05 & -8.84 & BLVC	 & 	37.52 & 66.08 & -7.73 & HVC	 & 	44.47 & 58.26 & -6.67 & HVC\\
    	&	& -40.99 & 64.09 & -8.69 & HVC	 & 	-48.36 & 117.16 & -7.58 & HVC	 & 	-47.17 & 121.59 & -6.46 & HVC\\
    	&	&  &  &  & 	 & 	-126.93 & 59.35 & -7.84 & HVC	 & 	-124.59 & 60.91 & -6.87 & HVC\\
\end{tabular}
    \caption{Centroid velocities, FWHM, luminosity and type of the individual components obtained by fitting the [OI] profiles from the MHD wind models. The fits to the [SII] profiles are listed in table \ref{tab:fits_mhd_pt2}.}
    \label{tab:fits_mhd_pt1}
\end{table*}

\begin{table*}
    \centering
    \begin{tabular}{crrrcrrrcrrrcr}
    & & \multicolumn{4}{c}{L$_{\mathrm{acc}}$ = $2.6\cdot10^{-2}$ L$_{\odot}$}
         & \multicolumn{4}{c}{L$_{\mathrm{acc}}$ = 0.31 L$_{\odot}$}
         & \multicolumn{4}{c}{L$_{\mathrm{acc}}$ = 1 L$_{\odot}$} \\ [1.5ex]
    \multirow{2}{*}{Line} & \multicolumn{1}{c}{i} 
        & \multicolumn{1}{c}{v$_\mathrm{c}$} & \multicolumn{1}{c}{FWHM} & \multirow{2}{*}{log\big($\mathrm{\tfrac{L_{comp}}{L_{\odot}}}$\big)} & \multicolumn{1}{c}{\multirow{2}{*}{Type}}
        & \multicolumn{1}{c}{v$_\mathrm{c}$} & \multicolumn{1}{c}{FWHM} & \multirow{2}{*}{log\big($\mathrm{\tfrac{L_{comp}}{L_{\odot}}}$\big)} & \multicolumn{1}{c}{\multirow{2}{*}{Type}}
        & \multicolumn{1}{c}{v$_\mathrm{c}$} & \multicolumn{1}{c}{FWHM} & \multirow{2}{*}{log\big($\mathrm{\tfrac{L_{comp}}{L_{\odot}}}$\big)} & \multicolumn{1}{c}{\multirow{2}{*}{Type}} \\
    & [\degree] 
        & \multicolumn{1}{c}{[km/s]} & \multicolumn{1}{c}{[km/s]} & & 
        & \multicolumn{1}{c}{[km/s]} & \multicolumn{1}{c}{[km/s]} & & 
        & \multicolumn{1}{c}{[km/s]} & \multicolumn{1}{c}{[km/s]} & &  \\
    \hline
    \hline
    \multirow{16}{*}{\rotatebox[origin=c]{90}{[SII] 4068}}
    	& \multirow{4}{*}{0}
    		& -21.05 & 21.34 & -6.27 & NLVC	 & 	-17.72 & 19.81 & -4.83 & NLVC	 & 	-17.89 & 20.04 & -4.06 & NLVC\\
    	&	& -48.65 & 46.80 & -5.81 & HVC	 & 	-39.73 & 43.62 & -4.91 & HVC	 & 	-40.29 & 44.19 & -4.12 & HVC\\
    	&	& -85.44 & 75.26 & -5.94 & HVC	 & 	-155.77 & 228.03 & -4.68 & HVC	 & 	-136.41 & 197.59 & -3.90 & HVC\\
    	&	& -174.01 & 161.97 & -6.05 & HVC	 & 	 &  &  & 	 & 	 &  &  & \\
    	\cline{2-14}
    	& \multirow{3}{*}{20}
    		& -25.86 & 42.00 & -5.95 & BLVC	 & 	-15.84 & 28.88 & -4.78 & NLVC	 & 	-16.27 & 27.72 & -4.01 & NLVC\\
    	&	& -62.40 & 71.02 & -5.75 & HVC	 & 	-44.62 & 56.37 & -4.91 & HVC	 & 	-43.87 & 53.76 & -4.13 & HVC\\
    	&	& -143.54 & 157.39 & -5.93 & HVC	 & 	-163.41 & 212.50 & -4.79 & HVC	 & 	-138.43 & 184.78 & -3.95 & HVC\\
    	\cline{2-14}
    	& \multirow{3}{*}{40}
    		& -4.15 & 40.76 & -6.10 & BLVC	 & 	3.89 & 20.35 & -5.29 & NLVC	 & 	2.07 & 19.57 & -4.52 & NLVC\\
    	&	& -45.81 & 78.99 & -5.75 & HVC	 & 	-23.42 & 54.52 & -4.69 & BLVC	 & 	-23.09 & 50.07 & -3.92 & BLVC\\
    	&	& -113.74 & 140.49 & -5.86 & HVC	 & 	-119.03 & 195.38 & -4.73 & HVC	 & 	-100.68 & 164.07 & -3.90 & HVC\\
    	\cline{2-14}
    	& \multirow{3}{*}{60}
    		& 11.69 & 27.99 & -6.78 & NLVC	 & 	13.96 & 23.38 & -5.68 & NLVC	 & 	12.60 & 21.48 & -4.78 & NLVC\\
    	&	& -57.01 & 127.93 & -5.89 & HVC	 & 	-24.78 & 49.94 & -5.37 & BLVC	 & 	-24.09 & 51.21 & -4.43 & BLVC\\
    	&	&  &  &  & 	 & 	-83.54 & 175.07 & -5.10 & HVC	 & 	-75.43 & 143.54 & -4.21 & HVC\\
    	\cline{2-14}
    	& \multirow{3}{*}{80}
    		& 24.21 & 50.73 & -6.32 & BLVC	 & 	20.11 & 43.24 & -5.48 & BLVC	 & 	24.70 & 28.45 & -4.83 & NLVC\\
    	&	& -47.08 & 83.46 & -6.09 & HVC	 & 	-24.12 & 26.70 & -5.89 & NLVC	 & 	-17.40 & 94.30 & -4.08 & BLVC\\
    	&	&  &  &  & 	 & 	-46.46 & 132.50 & -5.17 & HVC	 & 	-103.08 & 62.70 & -5.02 & HVC\\
    \hline
    \multirow{16}{*}{\rotatebox[origin=c]{90}{[SII] 6730}}
    	& \multirow{4}{*}{0}
    		& -16.31 & 18.52 & -6.97 & NLVC	 & 	-16.03 & 18.25 & -5.81 & NLVC	 & 	-16.36 & 18.39 & -5.19 & NLVC\\
    	&	& -37.82 & 41.31 & -6.57 & HVC	 & 	-36.06 & 38.71 & -5.93 & HVC	 & 	-36.36 & 38.41 & -5.29 & HVC\\
    	&	& -91.34 & 96.79 & -6.46 & HVC	 & 	-150.07 & 217.94 & -5.39 & HVC	 & 	-138.76 & 199.67 & -4.76 & HVC\\
    	&	& -204.03 & 139.76 & -6.37 & HVC	 & 	 &  &  & 	 & 	 &  &  & \\
    	\cline{2-14}
    	& \multirow{4}{*}{20}
    		& -15.84 & 24.50 & -6.83 & NLVC	 & 	-15.19 & 24.31 & -5.75 & NLVC	 & 	-15.42 & 24.06 & -5.14 & NLVC\\
    	&	& -38.83 & 44.38 & -6.60 & HVC	 & 	-40.44 & 48.58 & -5.95 & HVC	 & 	-39.53 & 46.30 & -5.31 & HVC\\
    	&	& -95.78 & 99.60 & -6.41 & HVC	 & 	-147.24 & 200.40 & -5.41 & HVC	 & 	-135.41 & 186.62 & -4.78 & HVC\\
    	&	& -201.17 & 120.08 & -6.45 & HVC	 & 	 &  &  & 	 & 	 &  &  & \\
    	\cline{2-14}
    	& \multirow{3}{*}{40}
    		& -10.52 & 32.86 & -6.70 & NLVC	 & 	-0.17 & 17.47 & -6.27 & NLVC	 & 	-0.69 & 17.81 & -5.63 & NLVC\\
    	&	& -41.15 & 59.14 & -6.59 & HVC	 & 	-20.63 & 40.26 & -5.69 & BLVC	 & 	-21.20 & 40.14 & -5.07 & BLVC\\
    	&	& -123.33 & 146.70 & -6.16 & HVC	 & 	-111.94 & 172.96 & -5.38 & HVC	 & 	-104.43 & 160.95 & -4.75 & HVC\\
    	\cline{2-14}
    	& \multirow{3}{*}{60}
    		& 9.67 & 19.81 & -7.03 & NLVC	 & 	10.39 & 19.93 & -6.14 & NLVC	 & 	10.25 & 19.84 & -5.50 & NLVC\\
    	&	& -20.36 & 51.87 & -6.77 & BLVC	 & 	-20.62 & 40.34 & -5.92 & BLVC	 & 	-20.74 & 41.72 & -5.26 & BLVC\\
    	&	& -73.87 & 116.27 & -6.12 & HVC	 & 	-75.30 & 127.96 & -5.43 & HVC	 & 	-72.33 & 121.22 & -4.80 & HVC\\
    	\cline{2-14}
    	& \multirow{2}{*}{80}
    		& 14.27 & 44.16 & -6.45 & BLVC	 & 	15.57 & 32.59 & -6.00 & NLVC	 & 	16.78 & 30.55 & -5.37 & NLVC\\
    	&	& -41.46 & 77.86 & -6.20 & HVC	 & 	-26.80 & 95.82 & -5.37 & BLVC	 & 	-25.53 & 92.99 & -4.73 & BLVC\\
\end{tabular}
    \caption{Centroid velocities, FWHM, luminosity and type of the individual components obtained by fitting the [SII] profiles from the MHD wind models. The fits to the [OI] profiles are listed in table \ref{tab:fits_mhd_pt1}.}
    \label{tab:fits_mhd_pt2}
\end{table*}

\begin{figure*}
    \centering
    \includegraphics[width=.3\textwidth]{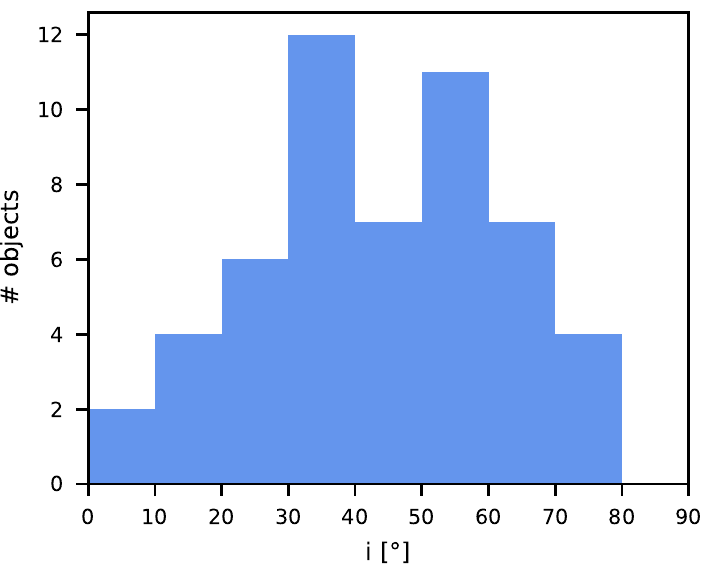}
    \caption{Number of objects in the sample of observations with a given inclination.}
    \label{fig:inclinations}
\end{figure*}\label{sec:appendix}


\bsp	
\label{lastpage}
\end{document}